\documentclass[letterpaper,11pt]{article}

\usepackage{jcappub}

\usepackage{tcolorbox}
\usepackage{latexsym}
\usepackage{amsfonts}
\usepackage{mathrsfs}
\usepackage{amssymb,amscd}
\usepackage[all]{xy}
\usepackage{amsmath}
\usepackage{amsthm}
\usepackage{fancyhdr}
\usepackage{fancybox}
\usepackage{xcolor}
\usepackage{hyperref}
\usepackage{verbatim}
\usepackage{makeidx}
\usepackage{indentfirst}
\usepackage[english]{babel}
\usepackage{graphicx}
\usepackage{listings}
\usepackage{subfigure}
\usepackage{multirow}
\usepackage{lscape}
\usepackage{physics}
\usepackage{amsmath}
\usepackage{framed}
\usepackage{bbold}
\usepackage{siunitx}
\usepackage{comment}
\usepackage{cases}
\usepackage{booktabs}

\usepackage{bm}
\numberwithin{equation}{section}

\usepackage{xparse}

\usepackage{hyperref}
\hypersetup{
    colorlinks=true,
    linkcolor=blue,
    filecolor=magenta,      
    urlcolor=cyan,
    pdftitle={Overleaf Example},
    pdfpagemode=FullScreen,
    }

\newcommand{\VEV}[1]{\left\langle #1 \right\rangle}
\newcommand{\threejsym}[6]{\left( {\begin{array}{ccc}
   #1 & #3 & #5\\
   #2 & #4 & #6\\ 
   \end{array}} \right)}

\newcommand{\tildeC}{\Tilde{C}}
\newcommand{\vecell}{\bm{\ell}}

\newcommand{\fsky}{f_{\text{sky}}}
\newcommand{\skewc}{\tilde{C}}
\newcommand{\ashot}{A_{\text{shot}}}

\def\d{{\rm d}}

\def\k{{\boldsymbol k}}

\def\q{{\boldsymbol q}}

\def\0{{\boldsymbol{0}}}

\def\G{{\cal G}}

\def\O{{\cal O}}

\def\nn{\nonumber\\}

\def\Ddelta{\delta_{\rm D}}
\def\Kdelta{\delta^{\rm K}}

\def\beq{\begin{equation}}
\def\eeq{\end{equation}}
\def\be{\begin{equation}}
\def\ee{\end{equation}}

\title{Skewing the CMB$\times$LSS: a Fast Method for Bispectrum Analysis}

\author{Priyesh Chakraborty$^{1}$, Shu-Fan Chen$^{2}$, and Cora Dvorkin$^{3}$} 
\affiliation{Department of Physics, Harvard University, 17 Oxford Street,
  Cambridge, MA 02138, USA}
\emailAdd{$^1$pchakraborty@g.harvard.edu,
  $^2$shufan\_chen@g.harvard.edu, 
  $^3$cdvorkin@g.harvard.edu,
 } 

\abstract{Upcoming cosmic microwave background (CMB) lensing measurements and tomographic galaxy surveys are expected to provide us with high-precision data sets in the coming years, thus paving the way for fruitful cross-correlation analyses. In this paper we study the information content of the weighted skew-spectrum, a nearly-optimal estimator of the angular bispectrum amplitude, as a means to extract non-Gaussian information on both bias and cosmological parameters from the bispectra of galaxies cross-correlated with CMB lensing, while gaining signiﬁcantly on speed. 
Our results show that for the combination of the {\it Planck} satellite and the Dark Energy Spectroscopic Instrument (DESI), the difference in the constraints on bias and cosmological parameters from the skew-spectrum and the bispectrum is at most 17\%.
We further compare and find agreement between our theoretical skew-spectra and those estimated from N-body simulations, for which it is important to include gravitational non-linearities beyond perturbation theory and the post-Born eﬀect for CMB lensing. We define an algorithm to apply the skew-spectrum estimator to the data and, as a preliminary step, we use the skew-spectra to constrain bias parameters and the amplitude of shot noise from the simulations through a Markov chain Monte Carlo likelihood analysis, finding that it may be possible to reach percent-level estimates for the linear bias parameter $b_1$.}

\begin{document}
\maketitle
\flushbottom

%%%%%%%%%%%%%%%%%%%%%%%%%%%%%%%%%%%
\section{Introduction}

With a variety of new cosmological data becoming available  this decade, we will be able to improve our understanding on many of the open questions that we are still facing in fundamental physics.
With galaxy surveys such as the Dark Energy Spectroscopic Instrument (DESI) \cite{DESI:2016fyo} and the {\it Vera Rubin} Observatory's Legacy Survey of Space and Time (LSST) \cite{Abell:2009aa} probing redshifts $z>1$, the amount of information accessible to us will be drastically growing. In addition to direct analyses of large-scale structure (LSS) surveys, it is useful to rely on cross-correlations with alternative tracers such as the cosmic microwave background (CMB). Through gravitational lensing, the observed CMB map can trace all intermediate matter between the surface of last scattering and the present day and, as such, is highly correlated with the intervening galaxy field. The $\textit{Planck}$ satellite \cite{Planck:2018vyg} has already detected CMB lensing to high statistical significance and upcoming CMB surveys such as CMB-Stage IV and the {\it Simons} Observatory will improve on this by an order of magnitude \cite{CMB-S4:2016ple,SimonsObservatory:2018koc}. A cross-correlation analysis is a useful complementary probe of the redshift evolution of LSS. Since the CMB lensing window peaks around $z\approx1-3$, correlation with galaxy samples in this range can provide, for example, constraints on the nature of dark energy \cite{Yu:2021vce} or tests of general relativity at the largest scales \cite{Pullen:2015vtb}. Moreover, it is a powerful tool to break degeneracies between cosmological parameters, such as the one between the matter fraction $\Omega_m$ and the amplitude of galaxy clustering measured on spheres of comoving radius $R=8/h$ Mpc, namely $\sigma_8$, and also bypass the cosmic variance limited by the survey volume~\cite{Seljak:2008xr}. Already at the power spectrum level, cross-correlation analyses are expected to boost constraints on new physics such as neutrino mass, primordial non-Gaussianity and shed light on existing cosmological tensions, such as the clustering of the LSS and the value of the Hubble parameter \cite{Schmittfull:2017ffw,Yu:2018tem,DES:2018miq,Zhang:2021eqp,Sgier:2021bzf,Sailer:2021yzm,White:2021yvw}.

Galaxy and CMB lensing maps are highly non-Gaussian due to gravitational instability and thus contain a wealth of information in higher-point correlation functions. However, working with higher-point statistics can be a complicated exercise since the number of configurations that are necessary to estimate them drastically increases. For example, while the power spectrum requires $\mathcal{O}(N^2)$ pairs, the bispectrum requires $\mathcal{O}(N^3)$ triangles, which can easily become computationally expensive as the number of observed modes $N$ grows. This issue can be circumvented in practice by averaging the bispectrum appropriately so that minimal information is lost, while effectively reducing the complexity of the analysis to that of a power spectrum, i.e. $\mathcal{O}(N\log{N})$ time. Such an averaging procedure can be performed in a number of ways, for example \cite{Cooray:2001ps} computes an un-weighted skew-spectrum of the CMB temperature field. The method we focus on in this paper is the \textit{bispectrum weighted} skew-spectrum (or simply the weighted skew-spectrum), which is derived from a minimum variance estimator of the bispectrum amplitude \cite{Babich:2004yc}. This approach was first introduced in \cite{Komatsu:2003iq} in order to expedite a non-Gaussian analysis of the CMB, and has subsequently been applied to the LSS in \cite{Schmittfull:2014tca,MoradinezhadDizgah:2019xun,Schmittfull:2020hoi,Munshi:2020ofi}. In this work, we extend the analysis to the cross-correlation between CMB lensing and LSS. Note that henceforth we will use the term \textit{skew-spectrum} to be concise, but our statistic should not be confused with that in \cite{Cooray:2001ps}, for example, where they do not apply a bispectrum weight. 

While this skew-spectrum was originally developed to detect a primordial bispectrum amplitude, it can be used to measure other amplitude parameters such as the galaxy bias parameters. As such, it is also an optimal statistic to capture physics that does not have its roots in primordial fluctuations. However, many cosmological parameters of interest do not, in fact, enter the bispectrum as an amplitude, e.g. the density $\Omega_{X}$ for species of matter, or the spectral tilt of the primordial power spectrum $n_s$. Despite this, we can attempt to constrain such parameters using the skew-spectrum and we investigate how the errors compare with those from the bispectra directly. In this work, we consider a $\nu\Lambda$CDM cosmology, where the neutrinos are parametrized by the total sum of neutrino masses $M_\nu=(m_{1}+m_{2}+m_{3})$, a precise measurement of which is instrumental in establishing the hierarchy of neutrino masses\footnote{Current experiments can only measure the difference of the squares of the masses, $\Delta m_{21}^2=m_2^2-m_1^2$ and $\abs{\Delta m^2}=\abs{m_3^2-(m_2^2+m_1^2)/2}$. Since the sign of $\Delta m^2$ cannot be determined, the hierarchy of masses is unknown \cite{Olive:2016}. The two contenders are the normal ($m_1\approx m_2 < m_3$) and inverted ($m_1<m_2\approx m_3$) hierarchies.}. Cross-correlations can be useful in constraining neutrino masses \cite{Dvorkin:2019jgs} already at the power spectrum level \cite{Yu:2018tem} and can be further improved by including the bispectrum \cite{Chen:2021vba}. 

As a preliminary step to an application to actual data sets, we apply our skew-spectrum to a simulation suite called {\tt MassiveNuS}~\cite{Liu:2017now}. In particular, the cross-correlation between CMB lensing and the LSS provides a valuable means of breaking degeneracies between cosmological amplitude parameters and galaxy bias, which is not possible using galaxy data alone. As such, we focus on using the cross-correlated skew-spectrum to constrain bias parameters for the {\tt MassiveNuS} halo catalogue, as well as the parameter $\ashot$, which measures the deviation from Poisson shot noise. This exercise has been performed in \cite{Schmittfull:2014tca} and \cite{Dai:2020adm} for the LSS alone with other simulation suites. 

This paper is organized as follows:
We start in Section~\ref{sec: skew spectrum} by recalling some relevant properties of our observables in harmonic space, define the skew-spectrum estimator and define the algorithm to be used in real data. In Section~\ref{sec: forecast}, we compare the bispectrum and the skew-spectrum through Fisher forecasts for CMB$\times$LSS, assuming DESI and the {\it Planck} satellite \cite{Planck:2018vyg} specifications. Finally, in Section~\ref{sec: sims}, we apply the skew-spectrum to the {\tt MassiveNuS} simulations as a proof-of-concept. We compute the halo and weak-lensing spectra, and covariance matrices from the simulated data. We further perform a Markov chain Monte Carlo (MCMC) likelihood analysis to infer constraints on halo bias parameters and $\ashot$. We conclude in Section~\ref{sec: conclusion}.

%%%%%%%%%%%%%%%%%%%%%%%%%%%%%%%%%%%
\section{Skew-Spectrum}\label{sec: skew spectrum}

To begin our discussion, we first note that any generic (local) quadratic field in harmonic space $D_{\ell m}$ is defined as a convolution of fields $\delta_{\ell m}$ with a kernel $K_{\ell_1\ell_2\ell_3}$:
\begin{align}
    D_{\ell m} = \sum_{\ell_1 m_1}\sum_{\ell_2 m_2} \threejsym{\ell_1}{m_1}{\ell_2}{m_2}{\ell}{-m}K_{\ell_1 \ell_2 \ell}\,\delta_{\ell_1 m_1}\delta_{\ell_2 m_2} \,,
\end{align}
where the Wigner $3\text{-}j$ symbol here is necessary to ensure that it is composed of a local product of density fields. 

Given the minimum variance (MV) estimator of the bispectrum amplitude, one can ask for the choice of kernels which minimizes loss of information when cross-correlated with a single density fluctuation from the full bispectrum. Namely, we are interested in the quantity,
\begin{align}
    \VEV{D_{\ell m}\delta^{*}_{\ell'm'}} &= \Kdelta_{\ell \ell'}\Kdelta_{mm'}\skewc_\ell\,,
\end{align}
where we call $\skewc_\ell$ the skew-spectrum and $\Kdelta$ is the Kronecker delta function. In the case of LSS, one such natural choice of kernels is the Legendre polynomials $\mathbf{P}_{\ell}$ for $\ell=0,\,1,\,2$\footnote{This is strictly true only for tree-level bispectra.}. This is because in galaxy bispectra, Fourier modes of density fluctuations couple with each other via the standard perturbation theory (SPT) kernels, which are composed of such orthogonal set of polynomials \cite{Schmittfull:2014tca,MoradinezhadDizgah:2019xun}. For example, $F_2$ is the lowest-order SPT kernel to enter the picture and is given by \cite{1980lssu.book.....P}
\begin{align}\label{eq:F2}
    F_{2}(\k_1,\k_2) &= \frac{17}{21} + \frac{1}{2}\left(\frac{k_1}{k_2}+\frac{k_2}{k_1}\right)\hat{\k}_{1}\cdot\hat{\k}_{2} + \frac{4}{21}\frac{3}{2}\left((\hat{\k}_{1}\cdot\hat{\k}_{2})^{2}-\frac{1}{3}\right)\nn
    &= \frac{17}{21}\mathbf{P}_{0}(\hat{\k}_1\cdot\hat{\k}_2) + \frac{1}{2}\left(\frac{k_1}{k_2}+\frac{k_2}{k_1}\right)\mathbf{P}_{1}(\hat{\k}_1\cdot\hat{\k}_2)+\frac{4}{21}\mathbf{P}_{2}(\hat{\k}_1\cdot\hat{\k}_2)\, .
\end{align}

In the following section, we will show a pedagogical derivation of the skew-spectrum in harmonic space and consider cross-correlations between two tracers, that is, galaxy surveys and CMB lensing in our case. 

%%%%%%%%%%%%%%%%%%%%%%%%%%%%%%%%%%%
\subsection{Power Spectrum and Bispectrum in Harmonic Space}\label{sub: maps and spectra}

We will begin by spelling out the power spectrum and bispectrum of the projected galaxy and CMB lensing convergence fields used in our analysis and in particular we will emphasize properties of the bispectrum that are crucial for a fast implementation of the skew-spectrum. 

The CMB weak lensing convergence field $\kappa(\hat{n})$, which is a projection of matter fluctuations from the surface of last scattering to the present, is given by~\cite{Lewis:2006fu,Bartelmann:1999yn}
\begin{align}
    \kappa(\hat{n}) &= \int_0^\infty \d\chi\, W_{\kappa}(\chi)\delta_m(\chi,\hat{n}),\\
    W_{\kappa}(\chi) &= \frac{3}{2}\Omega_{m0}H_0^2 \frac{1+z(\chi)}{H(\chi)}\frac{\chi(\chi_{s}-\chi)}{\chi_s}\theta(\chi_s-\chi),
\end{align}
where $\chi_s$ is the comoving distance to the last scattering surface, $H_0$ is the Hubble parameter today, $\Omega_{m0}$ is the fraction of matter density today, and $\theta(x)$ is the Heaviside theta function. Note that the matter density $\delta_m$ includes cold dark matter (CDM), baryons and massive neutrinos. Since gravitational lensing traces matter down to the lowest redshift and to small scales, the CMB lensing convergence field encodes non-Gaussian information.

On the other hand, galaxy samples populate a wide redshift coverage and have to be projected into several tomographic bins in order to be cross-correlated with CMB lensing. We define the projected galaxy density field as\footnote{Gravitational lensing can also modify the observed distribution of galaxies through magnification bias~\cite{Liu:2013yna,Hui:2007cu}. We do not include this effect here, but it should be taken into account in actual data analysis.}
\begin{align}
    \delta_g^{i}(\hat{n}) &= \int_0^\infty \d\chi\,  W_{g}^{i}(\chi)\delta_g(\chi,\hat{n})\,,\\
    W_{g}^{i}(\chi) &= \frac{1}{\Bar{n}_i}\frac{\d n_i}{\d\chi}\,,\quad\Bar{n}_i \equiv \int_0^\infty \d\chi\,\frac{\d n_i}{\d\chi}\,,
\end{align}
where $i$ labels the tomographic bin and $\d n_i/\d\chi$ is the number density of galaxies determined by the specification of the survey.

We are interested in the auto- and cross-correlated power spectra and bispectra of these two fields. Denoting $\{a,\,b,\,c\}$  $\in\{\kappa\,,g\}$, the power spectrum can be written as
\begin{align}
    \langle\delta^a_{\ell m}\delta_{\ell' m'}^{b*}\rangle &= \Kdelta_{\ell \ell'}\Kdelta_{m m'}C_{\ell}^{ab}\,,
\end{align}
and
\begin{align}
    C_{\ell}^{ab} &= \int_{0}^{\infty} \d\chi\, W_a(\chi)\int_{0}^{\infty} \d\chi' W_b(\chi')I_{\ell}^{ab}(\chi,\chi')\,, \\
    I_{\ell}^{ab}(\chi,\chi') &\equiv 4\pi \int_{0}^{\infty} \d k\,k^{2} j_{\ell}(k\chi) j_{\ell}(k\chi')P^{ab}(\chi,\chi';k)\,,
\end{align}
 where $j_{\ell}$ is the spherical Bessel function. Note that according to our convention, $C_{\ell}^{\kappa g}$ contains $P^{\kappa g}$ which should be understood as $P^{mg}$. Similarly, the bispectrum is expressed as
\begin{align}\label{eq:geomfactor}
    \langle\delta^a_{\ell_1 m_1}\delta_{\ell_2 m_2}^{b}\delta_{\ell_3 m_3}^c\rangle &= \mathcal{G}_{\ell_1 \ell_2 \ell_3}^{m_1 m_2 m_3}b_{\ell_1 \ell_2 \ell_3}^{abc}\,,
\end{align}
 where $\mathcal{G}_{\ell_1\ell_2\ell_3}^{m_1m_2m_3}$ is the Gaunt integral defined as
\begin{align}\label{eq:hdef}
    \mathcal{G}_{\ell_1 \ell_2 \ell_3}^{m_1 m_2 m_3} &\equiv g_{\ell_1\ell_2\ell_3}\threejsym{\ell_1}{m_1}{\ell_2}{m_2}{\ell_3}{m_3}\,,\\
    g_{\ell_1\ell_2\ell_3} &\equiv \sqrt{\frac{(2\ell_1+1)(2\ell_2+1)(2\ell_3+1)}{4\pi}}\threejsym{\ell_1}{0}{\ell_2}{0}{\ell_3}{0}\,.
\label{eq:geomfactor2}
\end{align}
The bispectrum can be schematically organized as follows \cite{Lee:2020ebj,Chen:2021vba}:
\begin{align}
    b^{abc}_{\ell_1\ell_2\ell_3} &= \sum_{n_1 n_2 n_3}c_{n_1 n_2 n_3}^{abc}\int_{0}^{\infty} \d r\, r^2 \Tilde{I}^{ac}_{\ell_1}(r;n_1)\Tilde{I}^{bc}_{\ell_2}(r;n_2)\Tilde{I}^{c}_{\ell_3}(r;n_3)\,,
    \label{eq:bispectrum}
\end{align}
where the factorized terms $\tilde{I}_\ell$ are
\begin{align}
    \Tilde{I}^{ab}_{\ell}(r;n) &\equiv 4\pi \int_{0}^{\infty} \d\chi\, W_a(\chi)D_+(\chi)\int_{0}^{\infty} \d k\, k^{2+n}j_{\ell}(kr)j_{\ell}(kr) P^{ab}(k)\,, \\ 
    \Tilde{I}^{a}_{\ell}(r;n) &\equiv 4\pi \int_{0}^{\infty} \d\chi\, W_a(\chi)D_+^2(\chi)\int_{0}^{\infty} \d k\, k^{2+n}j_{\ell}(kr)j_{\ell}(kr)\,.
\end{align}
Here, $D_{+}$ is the growth function normalized at $z=0$ and the coefficients $c_{n_1n_2n_3}^{abc}$ are numeric coefficients from the bias expansion and the SPT kernels. Despite the highly oscillatory spherical Bessel functions in the integrals $I_{\ell}^{ab}$, $\tilde{I}^{ab}_\ell$ and $\tilde{I}^{a}_\ell$, they can be computed efficiently with the use of the Limber approximation~\cite{Limber:1954zz} and the FFTLog algorithm~\cite{Assassi:2017lea,Lee:2020ebj,Chen:2021vba,Hamilton:1999uv}. Note that this separable form of the bispectrum based on $\tilde{I}_{\ell}$ is crucial for a fast estimation of the skew-spectrum from actual data and we will see how it works in Section~\ref{sub: convolution}. To be complete, we note that in the presence of massive neutrinos, the effective fluid description formally breaks down \cite{Senatore:2017hyk}. However, it has been demonstrated in \cite{Hu:1997vi,Takada:2005si} that it is sufficient to account for the neutrino free-streaming by including a non-trivial scale dependence in the growth function $D_+$, which is set by the free-streaming scale~\cite{Lesgourgues:2006nd} 
\begin{align}
    k_{\text{fs}} &\approx 0.23 \left(\frac{M_{\nu}}{0.1\text{ eV}}\right)\left(\frac{2}{1+z} \frac{\Omega_{m0}}{0.23}\right)^{1/2} h\text{Mpc}^{-1}\, ,
\end{align} 
below which neutrinos do not contribute to the clustering of structures. Using the Limber approximation, we can still preserve the separable form in the presence of a scale-dependent growth function (modulo some reorganizations), while with the FFTLog algorithm, we can make use of the expansion $D_+(\chi,k)=\sum_p\omega_p(k)\chi^{p}$ to still maintain the desired form~\cite{Chen:2021vba}, where $p=0,1,2,...$, and $\omega_p(k)$ are the coefficients of the polynomial expansion of the growth factor at each $k$.

%%%%%%%%%%%%%%%%%%%%%%%%%%%%%%%%%%%
\subsection{Skew-Spectrum Estimator}\label{sub: skew spec estimator}

We will briefly review the derivation of the skew-spectrum in this section.
We follow a similar procedure as in \cite{MoradinezhadDizgah:2019xun,Schmittfull:2014tca}, but modified for more than one field and in harmonic space. 

For the moment, let us consider a single observed field $\delta_{\ell m}$. Under the regime of weak non-Gaussianity, the MV estimator for the amplitude $f_{\rm NL}$ of the bispectrum is given as \cite{Babich:2004yc}
\begin{align}
    \hat{f}_{\rm NL} &=\frac{\sigma^2}{(4\pi\fsky)^2} \sum_{\ell_1\ell_2\ell_3}g_{\ell_1\ell_2\ell_3}^2\frac{b_{\ell_1\ell_2\ell_3}^{\rm th}\hat{b}^{\rm obs}_{\ell_1\ell_2\ell_3}}{C_{\ell_1}C_{\ell_2}C_{\ell_3}} \nonumber \\
    &= \frac{\sigma^2}{(4\pi\fsky)^2}\sum_{\ell_1 m_1}\sum_{\ell_2 m_2}\sum_{\ell_3 m_3}\G_{\ell_1\ell_2\ell_3}^{m_1m_2m_3}\,b_{\ell_1\ell_2\ell_3}^{\rm th}\frac{\delta_{\ell_1 m_1}\delta_{\ell_2 m_2}\delta_{\ell_3 m_3}}{C_{\ell_1}C_{\ell_2}C_{\ell_3}}\,,
    \label{eq:MV}
\end{align}
where $\sigma^2$ is the inverse variance of the estimator, $b^{\rm th}_{\ell_1\ell_2\ell_3}$ is the theoretical template, and $\hat{b}^{\rm obs}_{\ell_1\ell_2\ell_3}$ is the observed bispectrum. While we denote the estimator as $\hat{f}_{\rm NL}$, it can be applied to estimate the bispectrum amplitude of any non-Gaussian field. For example, when applied to the halo bispectrum, it may be used to infer the halo bias parameters.

Now we want to recast equation~\ref{eq:MV} as a sum over estimators of two-point statistics $\hat{\tilde{C}}_{\ell}$. Schematically, we want
\begin{align}
    \hat{f}_{\rm NL} &\propto \sum_{\ell} \hat{\tildeC}_{\ell}\,.
\end{align}
It is straightforward to verify that this reorganization is possible by identifying the quadratic field:
\begin{align}
    D_{\ell m} &= (-1)^m\sum_{\ell_1 m_1}\sum_{\ell_2 m_2}\G_{\ell_1\ell_2\ell}^{m_1m_2(-m)}\,b^{\text{th}}_{\ell_1 \ell_2 \ell}\frac{\delta_{\ell_1 m_1}}{C_{\ell_1}}\frac{\delta_{\ell_2 m_2}}{C_{\ell_2}}\,.
\end{align}
The odd factor $(-1)^m$ is simply a choice of phase which allows the skew-spectrum estimator to assume the canonical form of an estimator of a two-point function. Thus, we see that the requisite kernel to obtain the quadratic field is simply the theoretical template. Upon cross-correlating this quadratic field with a filtered field we obtain
\begin{align}
    \VEV{D_{\ell m}\frac{\delta^*_{\ell' m'}}{C_{\ell'}}} &= \Kdelta_{\ell\ell'}\Kdelta_{mm'}\sum_{\ell_1 \ell_2}\frac{g_{\ell_1\ell_2\ell}^2}{2\ell+1}\frac{b_{\ell_1 \ell_2 \ell}^{\text{th}}b_{\ell_1\ell_2\ell}^{\rm obs}}{C_{\ell_1}C_{\ell_2}C_{\ell}} =\Kdelta_{\ell\ell'}\Kdelta_{mm'} \tilde{C}_{\ell},
\end{align}
and we call this statistic the skew-spectrum. Consequently, the estimator for the skew-spectrum can be directly given as
\begin{align}
    \hat{\tildeC}_{\ell} &= \frac{1}{2\ell+1} \sum_{m}D_{\ell m}\frac{\delta_{\ell m}^*}{C_{\ell}}\,.
\end{align}
For multiple fields, this statistic can be generalized straightforwardly. For simplicity, let us define $\delta_{\ell m}^a/C^a_{\ell} \equiv \Tilde{\delta}_{\ell m}^{a}$. The quadratic field then becomes
\begin{align}
    D^{ab}_{\ell m} &= (-1)^m\sum_{\ell_1 m_1}\sum_{\ell_2 m_2}\G_{\ell_1\ell_2\ell}^{m_1m_2(-m)}\,b_{\ell_1 \ell_2 \ell}^{aab,\rm th}\Tilde{\delta}^a_{\ell_1 m_1}\Tilde{\delta}^a_{\ell_2 m_2}\,,
\end{align}
where we always assume that the quadratic field comes from only one field. We may cross-correlate this quadratic field with another field to obtain the skew-spectrum, which is defined as follows:
\begin{align}
    \VEV{D_{\ell m}^{ab} \tilde{\delta}^{b*}_{\ell' m'}} &= \Kdelta_{\ell \ell'} \Kdelta_{m m'} \tildeC_{\ell}^{ab}
\end{align}
and
\begin{align}\label{eq:skewSpecNoDecomp}
    \tildeC_{\ell}^{ab} &= \sum_{\ell_1 \ell_2}\frac{g_{\ell_1\ell_2\ell}^2}{2\ell+1}\frac{b_{\ell_1 \ell_2 \ell}^{aab,\text{th}}b_{\ell_1 \ell_2 \ell}^{aab,\rm obs}}{C^{a}_{\ell_1}C^{a}_{\ell_2}C^{b}_{\ell}}\,.
\end{align}
This quantity can also be readily estimated using standard power spectrum techniques,
\begin{align}
    \hat{\tildeC}^{ab}_{\ell} &= \frac{1}{2\ell+1}\sum_m D^{ab}_{\ell m} \Tilde{\delta}^{b*}_{\ell m}\,.
\end{align}

%%%%%%%%%%%%%%%%%%%%%%%%%%%%%%%%%%%
\subsection{Legendre Polynomial Decomposition of Skew-Spectrum}\label{sec:Leg Pol Dec}

Motivated by \cite{MoradinezhadDizgah:2019xun,Schmittfull:2014tca}, we follow a decomposition of the skew-spectrum based on Legendre polynomials and we analyze each of the terms as separate observables. In this section, we provide its details and specify its analytical covariance. We will compare the full and decomposed skew-spectra and conclude that the decomposition in Legendre polynomials helps the agreement with constraints from the full bispectrum.

At leading order in perturbation theory, the matter bispectrum is written as 
\begin{align}
    B^{mmm}(\k_1,\k_2,\k_3) = 2F_2(\k_1,\k_2)P(k_1)P(k_2) + \text{2 perms.}\, ,
\end{align}
where the kernel $F_{2}$ is the SPT kernel defined in equation \ref{eq:F2}. Going from matter overdensities to galaxy overdensities through a perturbative bias expansion\footnote{We assume a simple co-evolution halo bias model here as described in Appendix~\ref{A: tracer}. This is a simplified scenario and in practice we need to consider an additional halo occupation distribution model~\cite{Berlind:2001xk} for the galaxy density field.}  (see~\cite{Desjacques:2016bnm} for a review), we can write down the galaxy bispectrum (decomposed into Legendre polynomials $\mathbf{P}_\ell$) as 
\begin{align}\label{eq:bispec kSpace}
    B^{ggg}(\k_1,\k_2,\k_3) &= 2\left(\frac{17}{21}b_1^3+b_1^2\frac{b_2}{2}-\frac{2}{3}b_1^2b_{\G_2}\right)\mathbf{P}_{0}(\hat{\k}_1\cdot\hat{\k}_2)P(k_1)P(k_2) \nn
    &+2b_1^3\frac{1}{2}\left(\frac{k_1}{k_2}+\frac{k_2}{k_1}\right)\mathbf{P}_{1}(\hat{\k}_1\cdot\hat{\k}_2)P(k_1)P(k_2) \nn
    &+2\left(\frac{4}{21}b_1^3+\frac{2}{3}b_1^2b_{\G_2}\right)\mathbf{P}_{2}(\hat{\k}_1\cdot\hat{\k}_2)P(k_1)P(k_2)\nn 
    &+\text{2 perms.}
\end{align}
We can do the same decomposition for the angular galaxy bispectrum by simply projecting,
\begin{align}
    b^{ggg}_{\ell_1\ell_2\ell_3} = \left(\frac{17}{21}b_1^3+b_1^2\frac{b_2}{2}-\frac{2}{3}b_1^2b_{\G_2}\right)h_{0,\ell_1\ell_2\ell_3}^{ggg} + b_1^3h_{1,\ell_1\ell_2\ell_3}^{ggg} + \left(\frac{4}{21}b_1^3+\frac{2}{3}b_1^2b_{\G_2}\right)h_{2,\ell_1\ell_2\ell_3}^{ggg}\,,
    \label{eq:bggg}
\end{align}
where each of the $h_{i,\ell_1\ell_2\ell_3}^{aab}$ terms comes from different Legendre polynomials:
\begin{align}
    h_{0,\ell_1\ell_2\ell_3}^{aab} &\rightarrow \mathbf{P}_{0}(\hat{\k}_1\cdot\hat{\k}_2)P(k_1)P(k_2) + \text{2 perms.}\nonumber\\
    h_{1,\ell_1\ell_2\ell_3}^{aab} &\rightarrow \frac{1}{2}\left(\frac{k_1}{k_2}+\frac{k_2}{k_1}\right)\mathbf{P}_{1}(\hat{\k}_1\cdot\hat{\k}_2)P(k_1)P(k_2) + \text{2 perms.}\nonumber\\
    h_{2,\ell_1\ell_2\ell_3}^{aab} &\rightarrow \mathbf{P}_{2}(\hat{\k}_1\cdot\hat{\k}_2)P(k_1)P(k_2) + \text{2 perms.}\,.
    \label{eq:his}
\end{align}
Since the redshift evolution of the bias parameters are all absorbed into $h_{i,\ell_1\ell_2\ell_3}^{aab}$ due to the projection, the biases here just enter as amplitude parameters. Following this terminology, we can also get the kernels for cross-correlated bispectra of $gg\kappa$, $\kappa\kappa g$ and $\kappa\kappa\kappa$ from the three Legendre polynomials. Thus the skew-spectrum can now be written as 
\begin{align}\label{eq:skewSpecDecomp}
    \tilde{C}^{ab,i}_{\ell} = \frac{1}{2\ell+1}\sum_{\ell_1\ell_2}g_{\ell_1\ell_2\ell}^{2}\frac{h_{i,\ell_1\ell_2\ell}^{aab}b_{\ell_1\ell_2\ell}^{aab,\rm obs}}{C_{\ell_1}^{a}C_{\ell_2}^{a}C_{\ell}^{b}}\,,
\end{align}
where $g_{\ell_1\ell_2\ell}$ is the geometric factor defined in equation~\ref{eq:geomfactor2}. There is a difference between these kernels and the ones from \cite{MoradinezhadDizgah:2019xun,Schmittfull:2014tca}. They consider unsymmetrized kernels over $k_j$'s with $j=1,\,2,\,3$, while we impose symmetry in $h_{i,\ell_1\ell_2\ell_3}^{aab}$ over $\ell_j$'s.

Assuming a Gaussian covariance (i.e., we only consider the disconnected pieces), we can write down the diagonal and off-diagonal terms for the covariance matrix as
\begin{align}
    \langle\hat{\tilde{C}}^{ab,i}_\ell \hat{\tilde{C}}^{a'b',j}_{\ell'}\rangle_{\ell=\ell'} &=\frac{1}{2\ell+1}\frac{C_{\ell}^{D^{ab,i},D^{a'b',j}}C_{\ell}^{bb'}}{C_{\ell}^{b} C_{\ell}^{b'}}\nonumber\\
    &\quad+\frac{4}{(2\ell+1)^{2}}\frac{C_{\ell}^{ba'}C_{\ell}^{ab'}}{C_{\ell}^{b} C_{\ell}^{a'}C_{\ell}^{a} C_{\ell}^{b'}}\sum_{\ell_1}g_{\ell_1\ell\ell}^2 \frac{C_{\ell_1}^{aa'}}{C_{\ell_1}^{a}C_{\ell_1}^{a'}}h^{aab}_{i,\ell_1\ell\ell}h_{j,\ell_1\ell\ell}^{a'a'b'}\, \label{cov:1}, \\
    \langle\hat{\tilde{C}}^{ab, i}_{\ell} \hat{\tilde{C}}^{a'b', j}_{\ell'}\rangle_{\ell\neq\ell'} &= \frac{4}{(2\ell+1)(2\ell'+1)}\frac{C_{\ell}^{ba'}C_{\ell'}^{ab'}}{C_{\ell}^{b} C_{\ell}^{a'}C_{\ell'}^{a} C_{\ell'}^{b'}}\sum_{\ell_1}g_{\ell_1\ell\ell'}^{2} \frac{C_{\ell_1}^{aa'}}{C_{\ell_1}^{a}C_{\ell_1}^{a'}}h^{aab}_{i,\ell_1\ell'\ell}h_{j,\ell_1\ell\ell'}^{a'a'b'}\,\label{cov:2} ,
\end{align}
where
\begin{align}
    C^{D^{ab,i},D^{a'b',j}}_{\ell} &= \frac{2}{2\ell+1}\sum_{\ell_1 \ell_2}g_{\ell_1\ell_2\ell}^{2} \frac{C_{\ell_1}^{aa'}C_{\ell_2}^{aa'}}{C_{\ell_1}^{a} C_{\ell_2}^{a}C_{\ell_1}^{a'} C_{\ell_2}^{a'}}h_{i,\ell_1\ell_2\ell}^{aab}h_{j,\ell_1\ell_2\ell}^{a'a'b'} 
    \label{cov:DD}.
\end{align}

With the above equations in hand, we make a preliminary forecast to monitor the performance using only the first tomographic bin of DESI ELG, $z=[0.65,\,0.9]$. To begin, we vary only galaxy bias parameters and $A_{\rm shot}$~\cite{Dai:2020adm,Schmittfull:2014tca}, which measures the deviation from Poisson shot noise (see Section~\ref{sec: forecast} and Appendix~\ref{A: tracer} \& \ref{A: survey specs} for the full details). 

\begin{figure}[!h]
    \centering
    \includegraphics[scale=1.]{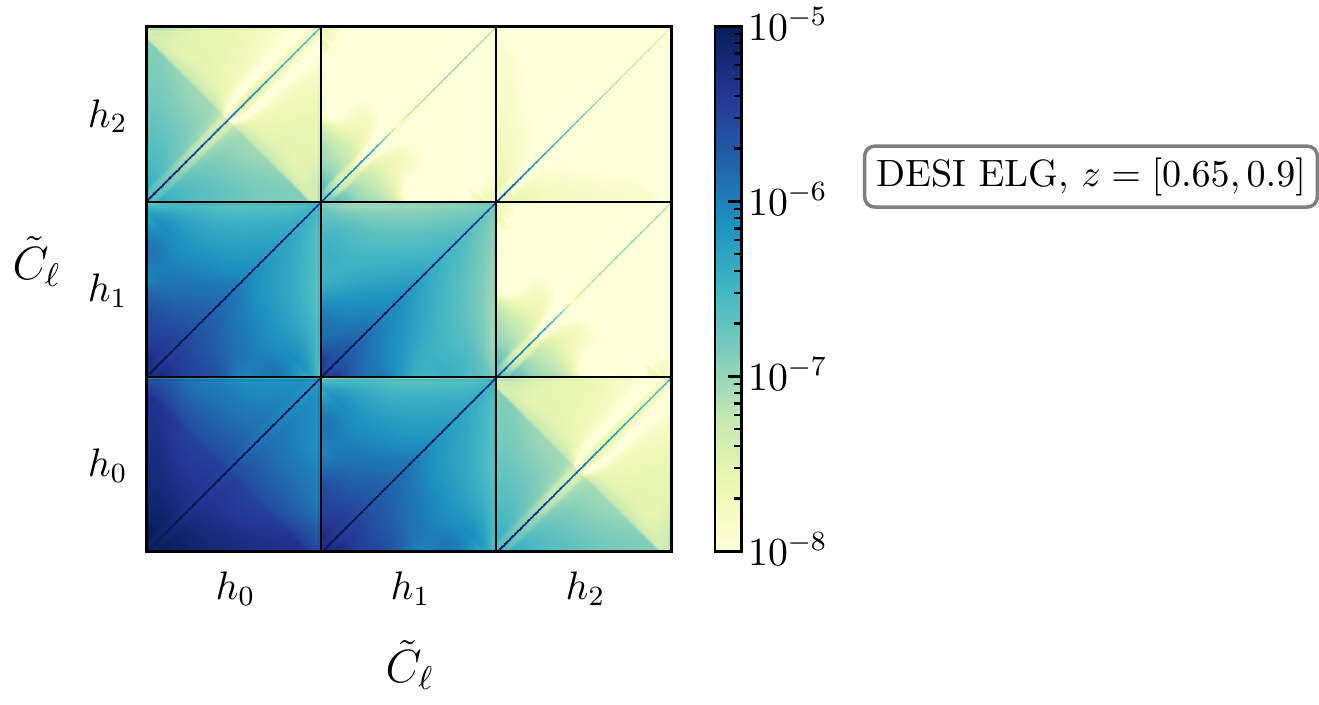}
    \caption{Covariance matrix of the skew-spectrum with decomposition into Legendre polynomials for the first tomographic bin of DESI ELG, $z=[0.65,\,0.9]$. The kernels $h_i$ (see equation \ref{eq:his}) here correspond to the different Legendre polynomials, as discussed in Section~\ref{sec:Leg Pol Dec}. The range of multipoles is $\ell\in[20,\,305]$ and is in increasing order within each submatrix.}
    \label{fig:CovarianceSep}
\end{figure}

In Figure~\ref{fig:CovarianceSep}, we show the covariance matrix of the decomposed skew-spectrum. Even though the diagonal components contribute the most in the covariance matrix, simply neglecting off-diagonal pieces can overestimate the constraints on parameters of interests (see~\cite{MoradinezhadDizgah:2019xun}). Therefore, in all the forecasts shown henceforth we include the off-diagonal terms as well.

\begin{figure}[!h]
    \centering
    \includegraphics[scale=0.8]{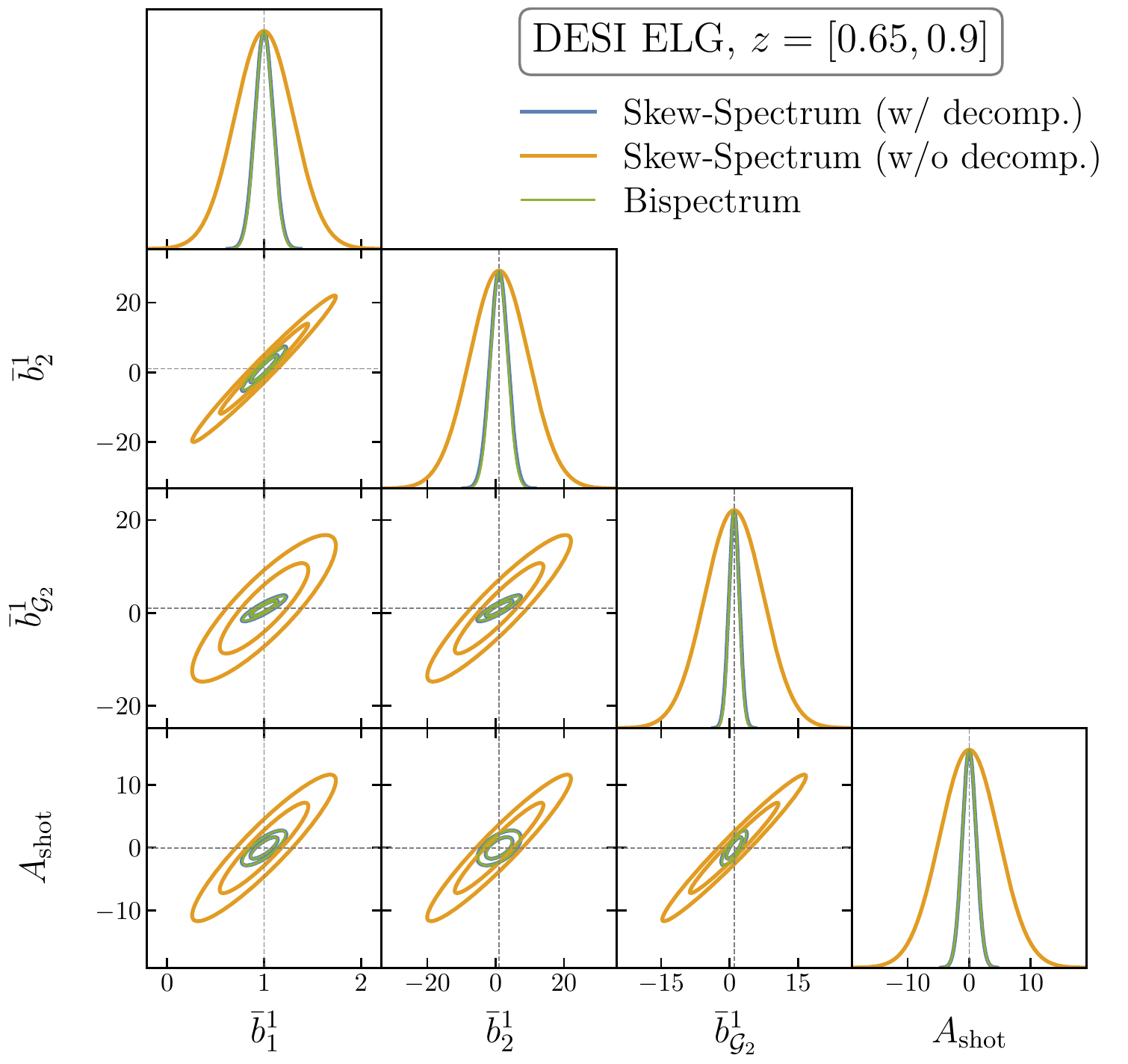}
    \caption{Forecasted $1\sigma$ and $2\sigma$ constraints on the bias parameters and $\ashot$ for the first tomographic bin of DESI ELG, $z=[0.65,\,0.9]$. Here we compare the constraints with and without applying a kernel decomposition on the skew-spectrum. For the case with decomposition, we analyze each kernel term as a separable observable given by equation~\ref{eq:skewSpecDecomp}, while for the case without decomposition, equation~\ref{eq:skewSpecNoDecomp} is used. We vary only bias parameters and $\ashot$ and do not marginalize over the $\nu\Lambda$CDM parameters here.}
    \label{fig:contourbiasSep}
\end{figure}

We first perform a comparison of forecasts with and without the decomposed kernels in the skew-spectrum\footnote{To obtain the covariance matrix for the case without the decomposition, we may simply replace $h^{ggg}_{i,\ell_1\ell_2\ell_3}$ in equations~\ref{cov:1}-\ref{cov:DD} by the leading-order theoretical $b^{ggg}_{\ell_1\ell_2\ell_3}$ in equation~\ref{eq:bggg}.} (see Figure~\ref{fig:contourbiasSep}). We can see that with the decomposition in Legendre polynomials, constraints from the skew-spectrum agree well with that from the bispectrum within $7\%$. On the other hand, skew-spectra without decomposition show worse constraints and degeneracy directions of the contours are tilted. Therefore, hereafter we only consider the spectra with the decomposed kernels.

In order to understand why we see an improvement when the skew-spectra are decomposed into Legendre polynomials, let us consider an example in 3D $k$-space for simplicity. As shown in equation~\ref{eq:bispec kSpace}, we can decompose the bispectrum as
\begin{align}\label{eq: bispecMultipole}
  B(k_1,k_2,\mu) = \sum_{\ell} B_{\ell}(k_1,k_2)P_{\ell}(\mu), 
\end{align}
with $\mu\equiv \hat{\k}_1\cdot\hat{\k}_2$. One thing to note here is that the relationship between $B(k_1,k_2,\mu)$ and $B_{\ell}(k_1,k_2)$ is invertible since the Legendre polynomials $P_{\ell}(\mu)$ form an orthonormal and complete basis. However, in the case of the skew-spectrum $\tilde{P}(k)$,
\begin{align}
    \tilde{P}(k) = \int \frac{\d^3q}{(2\pi)^3}\frac{B^{\rm th}(\k-\q,\q,-\k)B^{\rm obs}(\k-\q,\q,-\k)}{P(|\k-\q|)P(q)P(k)}\,,
\end{align}
when we insert equation~\ref{eq: bispecMultipole} in the above expression we obtain, \begin{align}\label{eq: redshift space skew decomp}
    \tilde{P}(k) = \sum_{\ell} \tilde{P}_{\ell}(k)\,,
\end{align}
where $\tilde{P}_\ell(k)$ is the skew-spectrum obtained using the redshift-space Legendre polynomial kernels (see the RHS of equation~\ref{eq:his}). Note that this expression is different from equation~\ref{eq: bispecMultipole} above, namely that there is no decomposition of the skew-spectrum itself occurring in an orthonormal basis. This implies that equation~\ref{eq: redshift space skew decomp} is not invertible, unlike the case of the bispectrum. Nevertheless, we see that knowing the $\tilde{P}_\ell(k)$ terms still allows us to determine $\tilde{P}(k)$ exactly. This implies that the decomposed spectra contain at least as much information as the full skew-spectrum. The lack of invertibility of equation~\ref{eq: redshift space skew decomp} means that the decomposed spectra must contain more information.

Moreover, there is a physical reason behind the improvement we observe. Recall that in Ref. \cite{Schmittfull:2014tca} the Legendre polynomial kernels were used to construct the quadratic field. It was pointed out that the skew-spectra computed with each Legendre polynomial kernel were mostly sensitive to the combination of bias parameters appearing as their pre-factors in the tree-level bispectrum (equation~\ref{eq:bispec kSpace}), therefore improving the constraints on each bias parameter overall.

We then study what happens when, in addition, we marginalize over the seven $\nu\Lambda$CDM parameters with the same DESI redshift bin. 
We see that constraints on galaxy biases from skew-spectra now becomes at most 1.3 times worse than those from the bispectrum, but still preserve the degeneracy lines. For the rest of the parameters, the degradation factors vary between 1.2 and 1.6, which are slightly worse than the biases, but also preserve the degeneracy directions. As we will see later in Section~\ref{sec: forecast}, this can be further improved with the combination of more tomographic bins and cross-correlations with CMB lensing.

%%%%%%%%%%%%%%%%%%%%%%%%%%%%%%%%%%%
\subsection{Data Application Algorithm}\label{sub: convolution}

While the bispectrum of Fourier modes benefits from factorization due to momentum conservation, the same is not true for the angular bispectrum. Namely, it is possible to write the bispectrum in momentum space into a separable form,
\begin{align}
    B(k_1, k_2, k_3) &= \sum_i f_i(k_1)g_i(k_2)h_i(k_3)
\end{align}
for some functions $f_i$, $g_i$ and $h_i$, but not for the one in harmonic space because of the line-of-sight projection. This factorizability in momentum space is crucial to recast the quadratic field as a convolution and thus improve the computational speed of the skew-spectrum estimation using fast Fourier transform algorithms~\cite{MoradinezhadDizgah:2019xun,Schmittfull:2014tca}. In order to achieve the same for angular bispectra, recall that the kernel can be expressed as
\begin{align}
    h^{abc}_{\ell_1\ell_2\ell_3} &= \sum_{n_1 n_2 n_3}c^{abc}_{n_1 n_2 n_3}\int_{0}^{\infty} \d r\, r^2 \Tilde{I}^{ac}_{\ell_1}(r;n_1)\Tilde{I}^{bc}_{\ell_2}(r;n_2)\Tilde{I}^{c}_{\ell_3}(r;n_3)\,.
\end{align} 
Thus, we can organize the quadratic field as follows,
\begin{align}
    D_{\ell m}^{ab} &= (-1)^m\sum_{n_1 n_2 n_3}c^{aab}_{n_1 n_2 n_3}\nonumber\\
    &\qquad\times\int_{0}^{\infty} \d r\, r^2 \sum_{\ell_1 m_1}\sum_{\ell_2 m_2} \mathcal{G}_{\ell_1\ell_2\ell}^{m_1 m_2(-m)}(\Tilde{I}_{\ell_1}^{ab}(r,n_1)\Tilde{\delta}^a_{\ell_1 m_1})(\Tilde{I}^{ab}_{\ell_2}(r,n_2)\Tilde{\delta}^a_{\ell_2 m_2})\tilde{I}_{\ell}^{b}(r,n_3) \nonumber \\
    &= (-1)^m\sum_{n_1 n_2 n_3}c^{aab}_{n_1 n_2 n_3}\int_{0}^{\infty} \d r\, r^2 D_{\ell m}^{ab}(r; n_1, n_2)\tilde{I}_{\ell}^{b}(r,n_3)\,,
    \label{eq:convolution}
\end{align}
where
\begin{equation}
    D_{\ell m}^{ab}(r;n_1,n_2) = \int \d^2\hat{n} \, Y^*_{\ell m}(\hat{n}) \Delta^{ab}(\hat{n};r,n_1) \Delta^{ab}(\hat{n};r,n_2)\,,
\end{equation}
and $\Delta^{ab}$ is defined as the field in position space with harmonic coefficients being
\begin{equation}
    \Delta_{\ell m}^{ab}(r,n) = \tilde{I}_{\ell}^{ab}(r,n)\tilde{\delta}_{\ell m}^a\,.
\end{equation}
Similar to the skew-spectrum in Fourier space, in order to get $D_{\ell m}^{ab}$, instead of summing over all the spherical harmonic coefficients as in equation~\ref{eq:convolution}, we can directly multiply the two $\Delta^{ab}$ fields in position space. Finally, we only need to cross-correlate the quadratic field $D_{\ell m}^{ab}$ with $\tilde{\delta}^{b}_{\ell m}$ in harmonic space. We summarize the roadmap as follows: 
\vskip 10pt
\begin{tcolorbox}[title=Skew-Spectrum Estimation Algorithm]
\begin{enumerate}
    \item Obtain fields $\delta^a(\hat{n})$ and $\delta^b(\hat{n})$.
    \item Compute the filtered harmonic coefficients, $\tilde{\delta}_{\ell m}^a$ and $\tilde{\delta}_{\ell m}^b$.
    \item Pointwise product between $\tilde{\delta}_{\ell m}^a$ and $\tilde{I}_{\ell}^{ab}(r,n_i)$ to obtain $\Delta^{ab}_{\ell m}(r,n_i)$.
    \item Return to position space to obtain $\Delta^{ab}(\hat{n};r,n_i)$.
    \item Compute harmonic transform of $\Delta^{ab}(\hat{n};r,n_1)\Delta^{ab}(\hat{n};r,n_2)$.
    \item Product with $\tilde{I}^{b}_{\ell}(r,n_3)$ and integrate over $r$ to obtain the quadratic field.
    \item Cross-correlate with $\tilde{\delta}^b_{\ell m}$ to obtain $\hat{\tildeC}^{ab}_{\ell}$.
\end{enumerate}
\end{tcolorbox}

%%%%%%%%%%%%%%%%%%%%%%%%%%%%%%%%%%%
\section{Forecast}\label{sec: forecast}

In this section we will compare the constraining power between the bispectrum and the skew-spectrum through a Fisher information analysis~\cite{Tegmark:1997rp}, as the first step to demonstrate its utility before proceeding to apply it on simulations. As shown in the previous section, it is necessary to decompose the kernels appropriately so as to achieve the optimal agreement with the bispectrum. After outlining our methodology, we will show a forecast using DESI ELG and {\it Planck} specifications as our reference. The survey specifications are described in Appendix~\ref{A: survey specs}.

%%%%%%%%%%%%%%%%%%%%%%%%%%%%%%%%%%%
\subsection{Methodology}

We perform a Fisher forecast to study the information content of skew-spectra and full bispectra. The data vector for the skew-spectrum consists of the three kernels and cross-correlations with CMB lensing, so it can be expressed as
\begin{align}
    \mathbf{d} &=\{\tilde{C}_{\ell}^{g_jg_j,i},\,\tilde{C}_{\ell}^{g_j\kappa,i},\,\tilde{C}_{\ell}^{\kappa g_j,i},\,\tilde{C}_{\ell}^{\kappa\kappa,i}\}\,,
\end{align}
where $i$ indicates the order of Legendre polynomials and $j$ labels the tomographic bins. Note that the data vector should include all $i$, $j$ and $\ell$s at the same time because there are correlations between most of them. The theoretical covariance matrix, $\mathbf{C}$, is computed using equations \ref{cov:1} and \ref{cov:2}. Therefore, given parameters $\boldsymbol{\lambda}$ of interests, we can write down the Fisher matrix $\mathbf{F}^{\rm skew}$ as
\begin{align}
    F^{\rm skew}_{\alpha\beta} = \frac{\partial\mathbf{d}}{\partial\lambda_\alpha}\mathbf{C}^{-1}\frac{\partial\mathbf{d}^{\rm T}}{\partial\lambda_\beta}\,.    
\end{align}
For the bispectrum, we consider the correlations $g_jg_jg_j$, $g_jg_j\kappa$, $\kappa\kappa g_j$ and $\kappa\kappa\kappa$ to match the data vector of the skew-spectrum. Assuming only Gaussian contributions in the covariance, the Fisher matrix $\mathbf{F}^{\rm bisp}$ can be written as~\cite{Babich:2004yc,Yadav:2007rk}
\begin{align}
    F_{\alpha\beta}^{\rm bisp} = \frac{f_{\rm sky}}{6}\sum_{abc}\sum_{a'b'c'}\sum_{\ell_1\ell_2\ell_3}g_{\ell_1\ell_2\ell_3}^2\frac{\partial b_{\ell_1\ell_2\ell_3}^{abc}}{\partial\lambda_\alpha}\left(\mathbf{C}_{\ell_1}'^{-1}\right)_{aa'}\left(\mathbf{C}_{\ell_2}'^{-1}\right)_{bb'}\left(\mathbf{C}_{\ell_3}'^{-1}\right)_{cc'}\frac{\partial b_{\ell_1\ell_2\ell_3}^{a'b'c'}}{\partial\lambda_\beta}\,,
\end{align}
where $\left(C_{\ell}'\right)_{aa'} = C_{\ell}^{aa'} + \Kdelta_{aa'}N_{\ell}^{aa'}$, $N_{\ell}$ is the noise power spectrum, and $g_{\ell_1\ell_2\ell_3}$ is the geometric factor defined in equation \ref{eq:geomfactor2}. With the Fisher matrix in hand, we can obtain lower bounds of the errors for each $\lambda_\alpha$ via the Cram\'er-Rao inequality: 
\begin{align}
    {\rm Var}(\lambda_{\alpha})\geq \left(\mathbf{F}^{-1}\right)_{\alpha\alpha}\,.
\end{align}

We use the best-fit {\it Planck} 2018 cosmological parameters as our fiducial cosmology \cite{Planck:2018vyg} and set the total sum of the neutrino mass $M_{\nu}$ to be 60 meV. Besides the three galaxy bias parameters in each tomographic bin, we also include $A_{\rm shot}$ to measure the deviation from the Poisson shot noise. That is, the parameters considered in our analysis are 
\begin{align}
    \boldsymbol{\lambda} &=\{H_0,\,\omega_b,\,\omega_c,\,A_s,\,n_s,\,\tau,\,M_\nu\}\cup\bigcup\limits_{i=1}^{N_{\rm bin}}\{\bar{b}_1^{i},\,\bar{b}_2^{i},\,\bar{b}_{\G_2}^{i}\}\cup\{\ashot\}\,,
\end{align}
where $N_{\rm bin}=4$ and $i$ labels the tomographic bin. Also note that $\bar{b}^i_{\O}$ parametrizes the amplitude of the bias parameters by $\bar{b}^i_{\O} b^i_{\O}(z)$ with $b^i_{\O}(z)$ being the fiducial redshift evolution of bias parameters, so the fiducial values for $\bar{b}^i_{\O}$ are all set to be 1.  Further, in our modelling of the galaxy power and bispectrum, we will assume a non-Poisson shot noise. It was shown in \cite{Baldauf:2013hka, Gil-Marin:2014sta} that the Poisson prediction should receive corrections due to clustering and exclusion effects of halos, especially on large scales. We will model the deviation of shot noise from the Poisson prediction through a phenomenological parameter $\ashot$. We provide details of our model in Appendix~\ref{A: tracer}.

To model the non-linear matter power spectrum, we implement {\tt Halofit}~\cite{Smith:2002dz,Mead:2020vgs} using {\tt CAMB}. For the modelling of the observed bispectrum, we go beyond the discussion in Section~\ref{sec:Leg Pol Dec}. First, we include the fitting formula from simulations for the SPT kernel $F_2$ to account for the bispectrum beyond the linear regime as it is accurate for scales $k\lesssim0.4\,h$Mpc$^{-1}$ and $z<1.5$. In this fitting formula, $F_{2}$ is modified as~\cite{Gil-Marin:2011jtv}
\begin{align}\label{eq:fitF2}
    F_{2}^{\rm fit}(\boldsymbol{k}_1,\boldsymbol{k}_2) &= \frac{5}{7}\tilde{a}(k_1)\tilde{a}(k_2) + \frac{1}{2}\left(\hat{\boldsymbol{k}}_1\cdot\hat{\boldsymbol{k}}_2\right)\left(\frac{k_1}{k_2}+\frac{k_2}{k_1}\right)\tilde{b}(k_1)\tilde{b}(k_2)\nonumber\\
    &+\frac{2}{7}\left(\hat{\boldsymbol{k}}_1\cdot\hat{\boldsymbol{k}}_2\right)^2\tilde{c}(k_1)\tilde{c}(k_2),
\end{align}
where functions $\tilde{a}$, $\tilde{b}$ and $\tilde{c}$ are determined by interpolation of simulations\footnote{ While more precise fitting functions for the bispectrum are available \cite{Takahashi:2019hth}, we choose to use this one for simplicity.}. We also include post-Born corrections for all bispectra that include CMB lensing, an effect which has been shown to be important \cite{Pratten:2016dsm,Fabbian:2019tik}. The inclusion of both the non-linear fitting function and the post-Born corrections to the theoretical bispectrum are needed to achieve a better agreement of the theoretical matter and weak lensing skew-spectra with those estimated from simulations (as we will show in Section~\ref{sec: sims}).

Even though we apply the fitting formula in the SPT kernel for our analysis, to be conservative we still restrict ourselves to the regime $k_{\rm max}\sim0.1$ $h$Mpc$^{-1}$. We set the range of multipoles to be $\ell=20-305$ for all tomographic bins. In order to facilitate the computation, we also set $\Delta\ell=2$ and rescale the Fisher matrix by $n_{\rm total}/n_{\rm sample}$ in the end, where $n_{\rm total}$ is the total number of possible non-vanishing configurations and $n_{\rm sample}$ is the number of configurations sampled. Furthermore, we do not include redshift-space distortions (RSD) since the effect was determined to affect the forecast only at the percent level \cite{Chen:2021vba}. We also note that the inclusion of RSD is more straightforward in harmonic space. In redshift space it is necessary to fix a LOS and therefore break isotropy, and this amounts to a modification of the skew-spectrum estimator \cite{Schmittfull:2020hoi} and the inclusion of a larger number of spectra. In harmonic space, it is only necessary to include extra terms in the theoretical bispectrum \cite{DiDio:2018unb}, but the estimator remains the same.

%%%%%%%%%%%%%%%%%%%%%%%%%%%%%%%%%%%
\subsection{CMB$\times$LSS Forecast}

\begin{figure}[!h]
    \centering
    \includegraphics[scale=0.5]{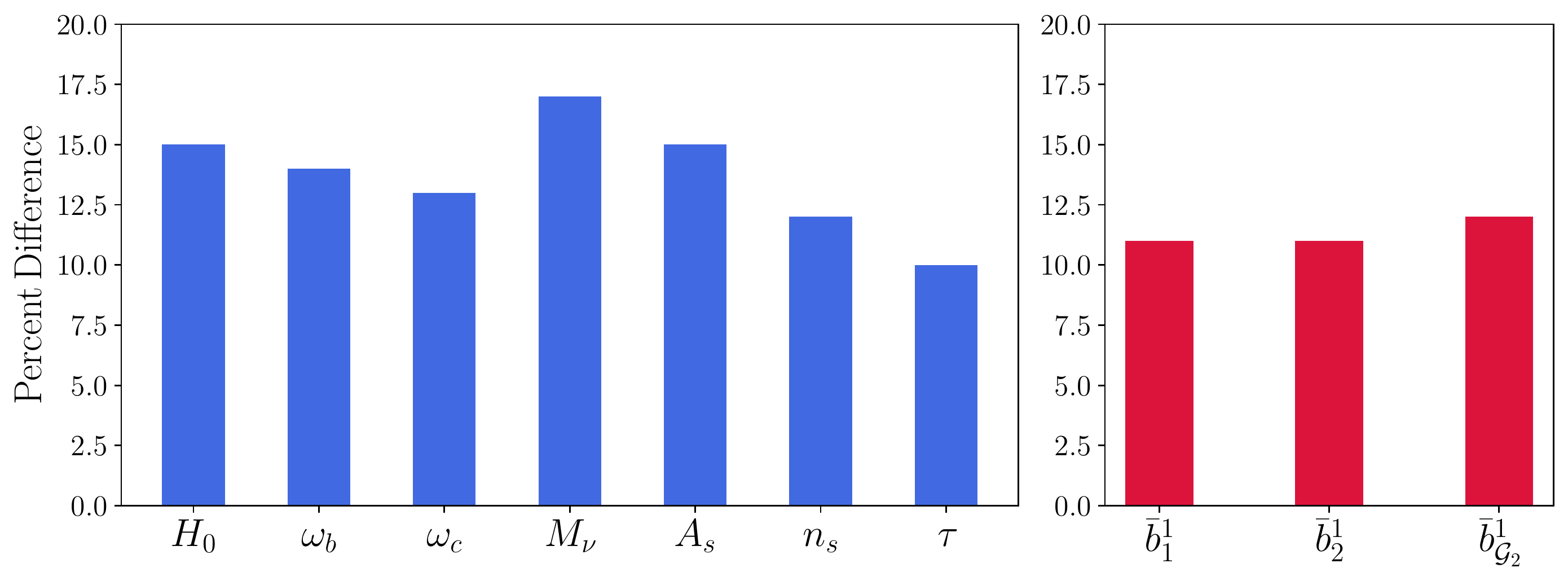}
    \caption{Relative difference in the $1\sigma$ forecasted constraints between DESI ELG $\times$ {\it Planck} lensing bispectrum and skew-spectrum. In the biases, we show only the last bin to be concise, but the rest show a relative difference of less than $12\%$. We also omit $\ashot$, which shows a relative difference of $50\%$.}
    \label{fig:DESI_CMBLensing}
\end{figure}

We present  in Figure~\ref{fig:DESI_CMBLensing} a comparison between the $1\sigma$ forecasted constraints from DESI ELG $\times$ {\it Planck} lensing bispectrum and skew-spectrum. We do not put any prior on $A_s$ in this case because the cross-correlation between galaxies and CMB lensing can already break the degeneracies between $A_s$ and galaxy biases. Compared to the simple forecast in Section~\ref{sec:Leg Pol Dec}, we find that adding the cross-correlation and including more tomographic bins enable a better agreement between the constraints from the skew-spectrum and the ones from the full bispectrum when we marginalize over the $\nu\Lambda$CDM parameters. 
Discrepancies for galaxy bias parameters display less than 12\% differences and at most 17\% for cosmological parameters.
The largest discrepancy is seen for the parameter $\ashot$ (50\%), due to the impact from cosmological parameters. This suggests that the skew-spectrum can be a robust estimator on non-amplitude parameters without direct estimation of the full bispectrum from the data. It is therefore also expected that the two approaches for $C_{\ell}+B_{\ell_1\ell_2\ell_3}$ and $C_{\ell}+\tilde{C}_{\ell}$ should be competitive with each other, but this is not the focus of this paper.

We note that from a joint analysis of CMB lensing and LSS we expect tight constraints mainly on $\omega_m$, $\sigma_8$, $M_\nu$, along with the galaxy bias parameters. However, we choose to vary the full cosmology in order to achieve a more complete sense of the non-Gaussian information content of the skew-spectra.

%%%%%%%%%%%%%%%%%%%%%%%%%%%%%%%%%%%
\section{Simulations}\label{sec: sims}

In preparation for the application of our formalism to real data, we test it on the {\tt MassiveNuS} simulation suite\footnote{\url{http://ColumbiaLensing.org}}~\cite{Liu:2017now}. These simulations contain $1024^3$ CDM particles with neutrino patches in a box of size $512$ $h^{-1}$Mpc and contain $67$ snapshots from $z=0$ up to $z=45$. We use the publicly available fiducial model of the simulation with $M_{\nu}=0.1$ eV. We also utilize the halo catalogue offered by {\tt MassiveNuS} obtained by a friends-of-friends-based algorithm. For selection and analysis of the halo catalogue, we use the publicly available code {\tt Halotools}\footnote{\url{https://halotools.readthedocs.io/en/latest/}}~\cite{Hearin:2016uxs}.

In order to obtain the projected halo number overdensity and the CMB lensing convergence field, we make use of the publicly available code {\tt LensTools}\footnote{\url{https://pypi.python.org/pypi/lenstools/}}\cite{Petri:2016zrq}. The lensing maps produced by {\tt LensTools} are obtained by ray-tracing through planes cut from the simulation box. We modify the code to generate the projected halo number overdensity field in order to cross-correlate with the lensing convergence field. We cut each simulation box into four planes from each of the three normal directions with thickness 126 $h^{-1}$Mpc at each sampled redshift of {\tt MassiveNuS}. Then we shoot the photons from $z=0$ back to the last scattering surface with {\tt LensTools} to generate 10240 CMB lensing convergence maps from one simulation box~\cite{Petri:2016wlu}. The sky coverage of the angular maps are of $3.5\times3.5$ deg$^{2}$ in size with 2048 pixels per side. The $\ell_{\rm min}$ is set to be 150 with bin size $\Delta \ell = 300$. To avoid non-linearities at small scales, we put a sharp cutoff at $\ell=1350$. 

To proceed with the cross-correlation, we select a halo sample with minimum mass $1.36\times 10^{12}\,h^{-1}M_{\odot}$, which corresponds to approximately $3.5\times10^5$ halos in the whole simulation box. Our halos populate a single plane at $z=1.04$. Thus in order to be consistent in our model, we use a top-hat window function centered at $z=1.04$. Since the simulation data is at non-linear scales and further, the ray-traced lensing convergence maps go beyond the Born approximation, we also apply the same model for the observed bispectra discussed in Section~\ref{sec: forecast}, which includes the fitting formula $F_{2}^{\rm fit}$ and post-Born corrections in CMB lensing. As a sanity check for our modifications of {\tt LensTools}, we compare the theoretical predictions of cross-correlated power spectra and bispectra between projected matter density and CMB lensing convergence and find agreement with the simulation\footnote{We use binned power spectrum and bispectrum estimators~\cite{Coulton:2018ebd,Bucher:2009nm,Bucher:2015ura} to compute the spectra from simulations.}.

The sky coverage of the simulated maps is small, so we can safely use the flat-sky approximation. We first rewrite our skew-spectra in flat-sky coordinates in Section~\ref{sec:flatsky} and compare with the full-sky case in the forecasted errors (including cosmological parameters), finding agreement up to $10\%$ in all of the parameters. Note that for this comparison, we assume $f_{\rm sky}=1$. Further, we do not use the Legendre polynomial kernels and use the full bispectrum kernel instead, with the expectation that the agreement will hold when considering the decomposed kernels.
Finally, we estimate the halo biases and the parameter $\ashot$ from the simulations using an MCMC likelihood analysis. We present details of the bias model and shot noise in Appendix~\ref{A: tracer}. 

%%%%%%%%%%%%%%%%%%%%%%%%%%%%%%%%%%%
\subsection{Flat-Sky Approximation}\label{sec:flatsky}

\begin{figure}[!h]
    \centering
    \includegraphics[scale=0.5]{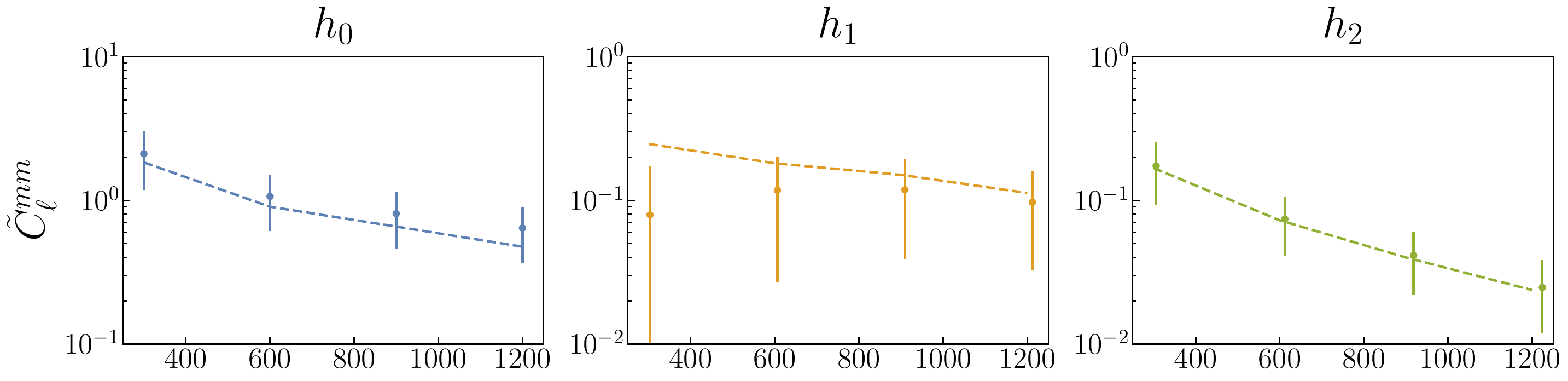}
    \includegraphics[scale=0.5]{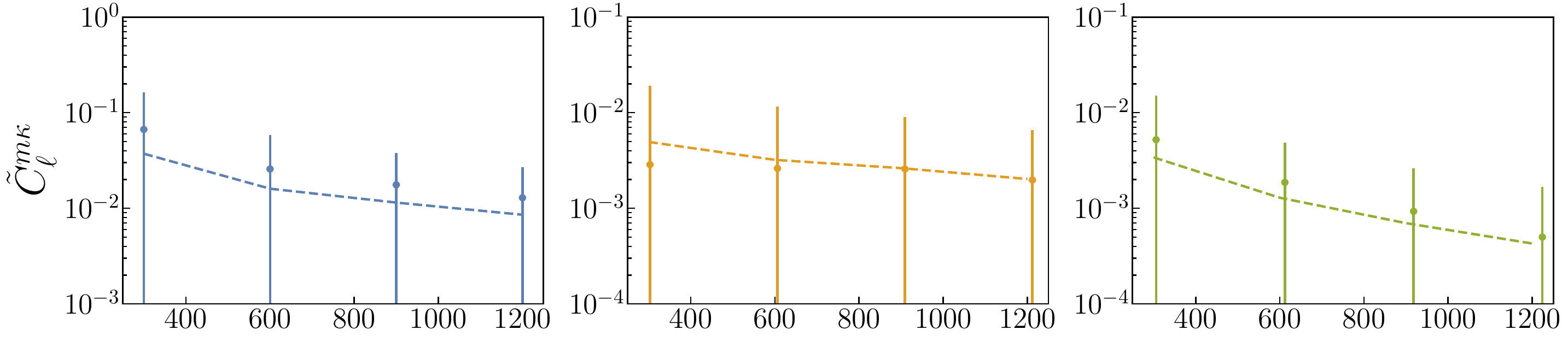}
    \includegraphics[scale=0.5]{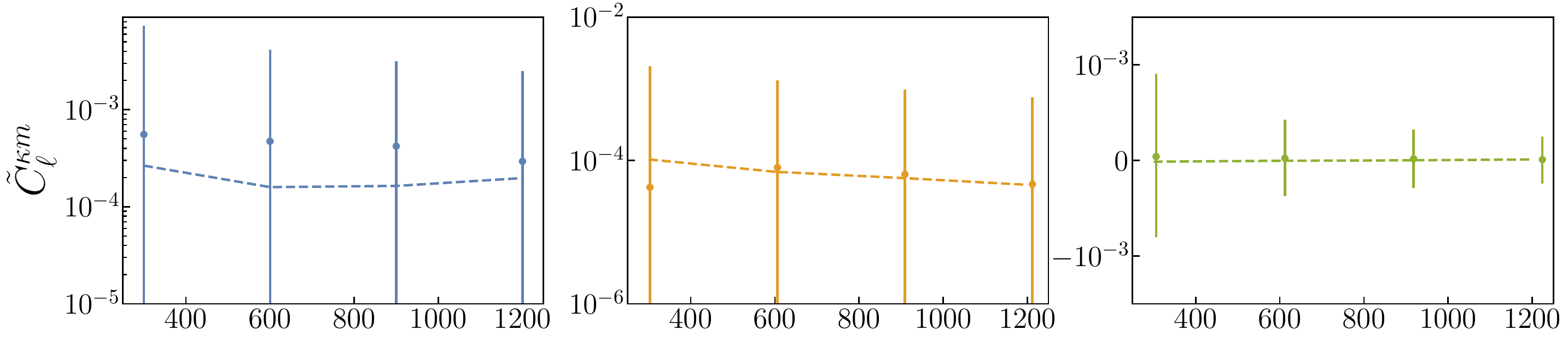}
    \includegraphics[scale=0.5]{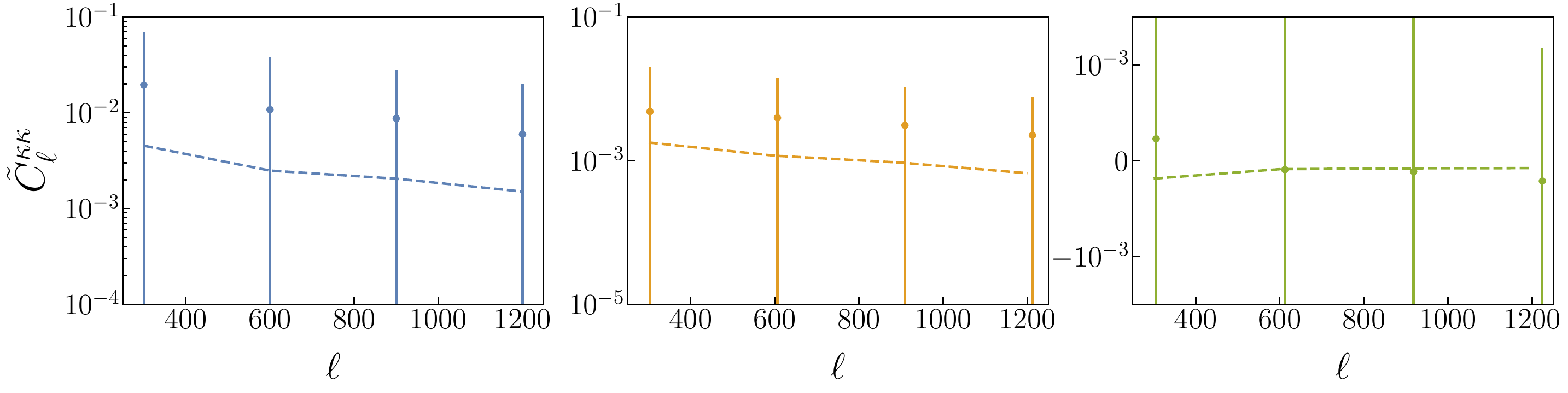}
    \caption{Comparison of the skew-spectra of CMB weak lensing convergence$\times$matter fields from theoretical predictions and simulations. The dashed lines show the theoretical predictions when using $F_{2}^{\rm fit}$ in equation~\ref{eq:fitF2} and the inclusion of post-Born corrections. The dots show the average value of the skew-spectra estimated from the simulation and error bars are the variance of them. We can see agreements within the error bars in most of the comparisons.}
    \label{fig:skewspectrumFlatSkySimulation}
\end{figure}

Since the sky coverage of our simulation data is small, it is sufficient to use the flat-sky approximation here, which allows gains in speed while minimizing loss of information. To proceed, we note the following mappings between spherical harmonic coefficients and flat-sky Fourier modes \cite{Hu:2000ee}:
\begin{align}
    \delta_{\ell m} &= \sqrt{\frac{2\ell+1}{4\pi}}i^m \int\frac{\d\varphi_\ell}{2\pi}e^{-im\varphi_\ell}\delta(\vecell)\,, \\
    \delta(\vecell) &= \sqrt{\frac{4\pi}{2\ell+1}}\sum_m i^{-m}\delta_{\ell m} e^{im\varphi_\ell}\,.
\end{align}
Using these mappings we obtain the quadratic filtered field,
\begin{align}\label{eq:flatquad}
    D_{i}^{ab}(\vecell) &= \int_{\vecell_1} \int_{\vecell_2} (2\pi)^2\Ddelta(\vecell-\vecell_1-\vecell_2)h^{aab}_{i}(\vecell_1,\vecell_2,\vecell)\delta^a(\vecell_1)\delta^a (\vecell_2)\,.
\end{align}
Here $h_{i}^{aab}$ is analogous to the convolution kernel in full sky. While the bispectrum above is strictly the three-point function of flat-sky modes, for scales under consideration it is identical within a few percent to the full-sky bispectrum. Thus, we can approximate the flat-sky theoretical bispectrum with the full-sky one:
\begin{align}
    \VEV{\delta^a(\vecell_1)\delta^b(\vecell_2)\delta^c(\vecell_3)} &= (2\pi)^2\Ddelta(\vecell_1+\vecell_2+\vecell_3)b^{abc}(\vecell_1,\vecell_2,\vecell_3) \nonumber \\
    &\approx(2\pi)^2\Ddelta(\vecell_1+\vecell_2+\vecell_3)b^{abc}_{\ell_1\ell_2\ell_3}\,.
\end{align}
The theoretical skew-spectrum then becomes
\begin{align}
    \tilde{C}^{ab}_{i,L} &= \frac{1}{M_{L}}\sum_{\boldsymbol{\ell}\in L}\int_{\boldsymbol{\ell}_1} \frac{h_{i}^{aab}(\boldsymbol{\ell}_1,\boldsymbol{\ell-\ell}_1,\boldsymbol{-\ell})b^{aab}(\boldsymbol{\ell}_1,\boldsymbol{\ell-\ell}_1,\boldsymbol{-\ell})}{C^{a}_{\ell_1} C^{a}_{|\boldsymbol{\ell-\ell}_1|}C^{b}_{\ell}}\,,
\end{align}
where $L$ labels the bin and $M_L$ is the number of modes in bin $L$. We can straightforwardly write the estimator as
\begin{equation}
    \hat{\tilde{C}}^{ab}_{i,L} = \frac{1}{M_L}\sum_{\boldsymbol{\ell}\in L} D^{ab}_{i}(\boldsymbol{\ell})\tilde{\delta}^{b}(-\boldsymbol{\ell})\,.
\end{equation}
We account for binning by replacing the angular integral in Fourier space as $\int \tfrac{\d\varphi_{\ell}}{2\pi}\rightarrow\tfrac{1}{M_L}\sum_{\vecell\in L}$. Note that we can still have a fast estimation of the skew-spectrum with the use of the FFTLog algorithm in equation~\ref{eq:flatquad}, where convolution becomes a product of pixels of two maps in position space (see Section~\ref{sub: convolution}). 
Finally, the theoretical covariance is
\begin{align}
    \langle \hat{\tilde{C}}^{ab}_{i,L},\,\hat{\tilde{C}}^{a'b'}_{j,L'}\rangle = \delta_{LL'}I_{ij,L}^{ab,a'b'} + \frac{1}{4\pi f_{\rm sky}}J_{ij,LL'}^{ab,a'b'}\,,
\end{align}
where $I_{ij,L}$ and $J_{ij,LL'}$ are
\begin{align}
    I_{ij,L}^{ab,a'b'} &\equiv \frac{2}{M_{L}^2}\sum_{\boldsymbol{\ell}\in L}\int_{\boldsymbol{\ell}_1}\frac{C^{bb'}_{\ell}}{C^b_{\ell}C^{b'}_{\ell}}\frac{C_{\ell_1}^{aa'}C_{|\boldsymbol{\ell-\ell}_1|}^{aa'}}{C_{\ell_1}^{a}C_{|\boldsymbol{\ell-\ell}_1|}^{a}C_{\ell_1}^{a'}C_{|\boldsymbol{\ell-\ell}_1|}^{a'}}h_i^{aab}(\boldsymbol{\ell}_1,\boldsymbol{\ell-\ell}_1,-\boldsymbol{\ell})h_j^{a'a'b'}(\boldsymbol{\ell}_1,\boldsymbol{\ell-\ell}_1,-\boldsymbol{\ell})\,, \\
    J_{ij,LL'}^{ab,a'b'} &\equiv \frac{4}{M_{LL'}}\sum_{\boldsymbol{\ell}\in L,\,\boldsymbol{\ell}'\in L'}\frac{C_{\ell'}^{ab'}C_{\ell}^{ba'}}{C_{\ell'}^aC_{\ell}^bC_{\ell}^{a'}C_{\ell'}^{b'}}\frac{C_{|\boldsymbol{\ell-\ell}'|}^{aa'}}{C_{|\boldsymbol{\ell-\ell}'|}^aC_{|\boldsymbol{\ell-\ell}'|}^{a'}}h_i^{aab}(\boldsymbol{\ell-\ell}',\boldsymbol{\ell}',-\boldsymbol{\ell})h_j^{a'a'b'}(\boldsymbol{\ell-\ell}',-\boldsymbol{\ell},\boldsymbol{\ell}').
\end{align}
Here, $M_{LL'}$ denotes the number of modes from the combination of bins $L$ and $L'$. We show a comparison of the skew-spectrum between theoretical predictions and simulations in Figure~\ref{fig:skewspectrumFlatSkySimulation} and observe a general agreement within error bars. A comparison between the theoretical covariance and the estimated covariance from simulations is shown in Appendix~\ref{A:Covariance}, where we also see a general agreement. However, it also suggests a more accurate modeling of small-scale spectra is required since the covariance coming from simulations shows a larger amplitude for the $\kappa g$, $g\kappa$ and $\kappa\kappa$ spectra. This implies that the error bars in our Fisher analysis are likely underestimated.

%%%%%%%%%%%%%%%%%%%%%%%%%%%%%%%%%%%
\subsection{Bias Estimation}

\begin{figure}[!h]
    \centering
    \includegraphics[width=0.7\textwidth]{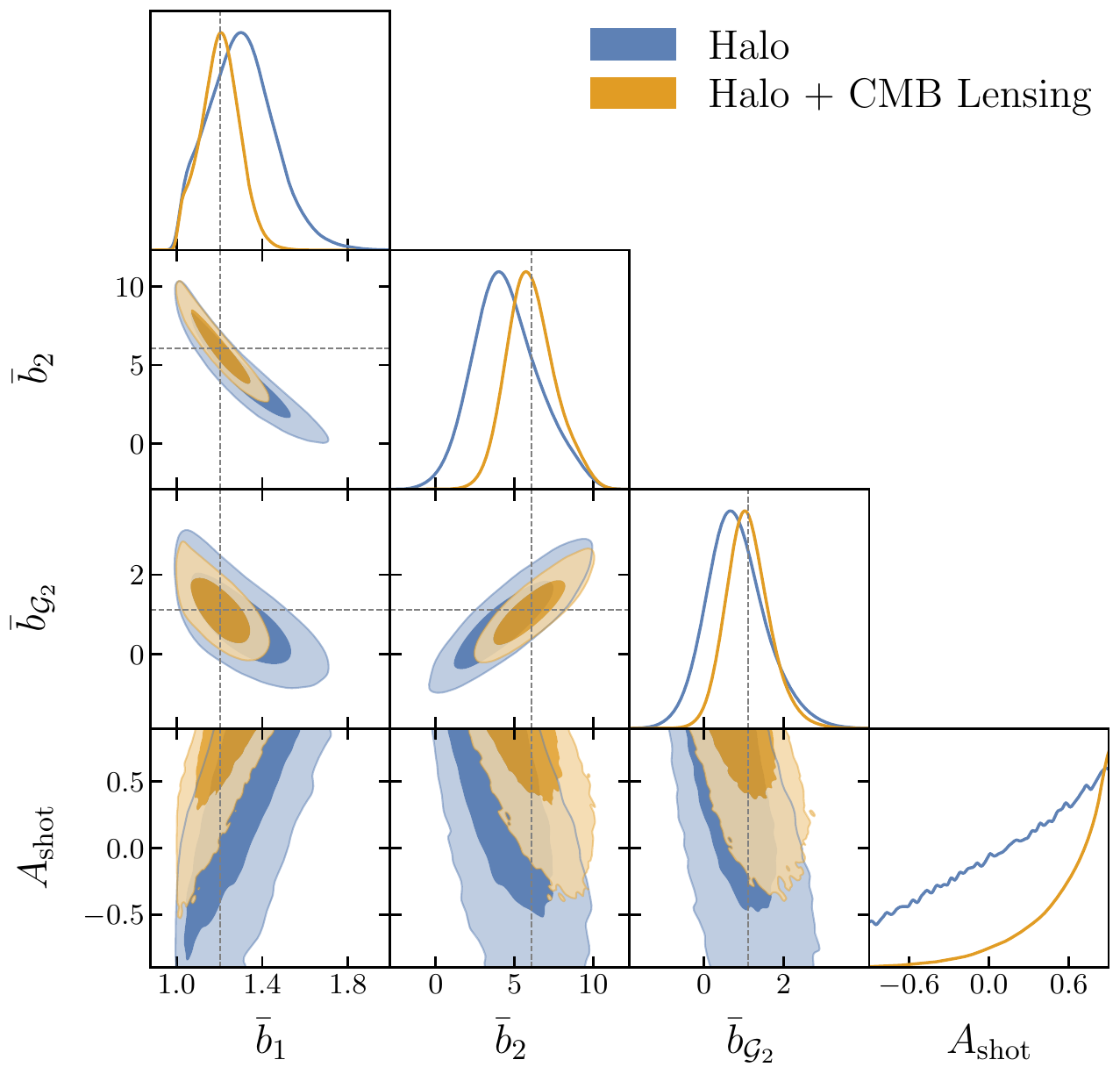}
    \caption{Posterior distributions of bias parameters and $A_{\text{shot}}$ at $z=1.04$ in the {\tt MassiveNuS} simulations from an MCMC likelihood analysis on the skew-spectra for $\kappa\kappa\kappa$, $hhh$, $\kappa$$hh$, and $h$$\kappa\kappa$ (brown contours). The dark and light shaded regions correspond to the $1\sigma$ and $2\sigma$ levels and the dashed lines show the means of the posteriors. We also show the case for $hhh$ skew-spectra only (blue contours) to illustrate the improvement when combining with CMB lensing. Note that we expect the contours to change when cosmological parameters are varied in the analysis.}
    \label{fig:bias_posterior}
\end{figure}

\begin{figure}[!h]
    \centering
    \includegraphics[width=1\linewidth]{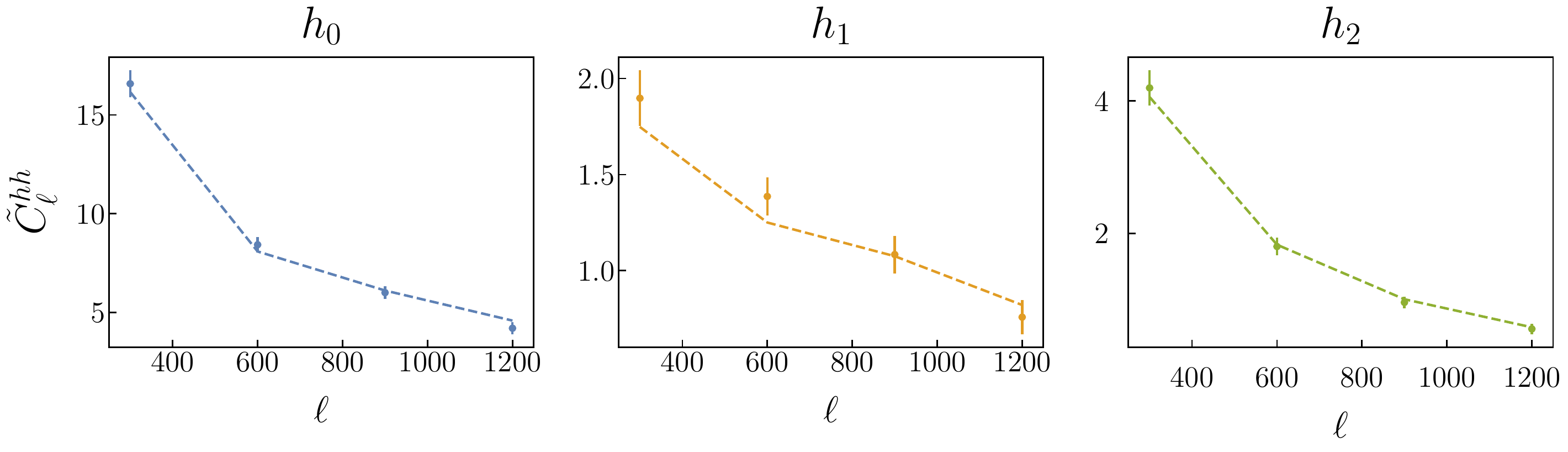}
    \caption{MCMC results for the halo-halo skew-spectra with the monopole, dipole and quadrupole kernels. The dashed lines shows the simulation data and the points show the theoretical skew-spectra using the mean values of the parameters from the marginalized posteriors. The error bars are computed from the analytical covariance matrix.}
    \label{fig:clhhMCMC}
\end{figure}

So far we have only shown the constraining power of the skew-spectra using Fisher analysis. In this section, we aim to estimate halo bias parameters with an MCMC as a practical application of our method. For this purpose, we use the publicly available Monte Carlo sampler {\tt emcee}\footnote{\url{https://emcee.readthedocs.io/en/stable/}}~\cite{2013PASP..125..306F} to estimate the posterior probability distributions of the bias parameters, $\bar{b}_1$, $\bar{b}_2$, and $\bar{b}_{\G_2}$, and $A_{\rm shot}$ using halo catalogues from {\tt MassiveNuS}. Since the redshift window is narrow, we assume that the bias parameters are constant over redshift when analyzing the simulations. In contrast to~\cite{Schmittfull:2014tca,Dai:2020adm}, which conduct a momentum-space analysis of halo-halo skew-spectra alone, we perform the analysis in harmonic space and include cross-correlation with the CMB lensing convergence field. The setup for the simulation and the selection for halos are the same as described earlier in this section. We assume a Gaussian likelihood and use the analytically computed covariance since we have only one available independent simulation at hand. We adopt non-informative uniform priors on all the four parameters, as shown in Table~\ref{tab:MCMCPrior}. Note that $\ashot$ is expected to lie within $[-1,\,1]$~\cite{Gil-Marin:2014sta} and our prior reflects this. We present our posterior distributions in Figure~\ref{fig:bias_posterior} and show the errors of each parameter in Table~\ref{tab:MCMCPrior}. We also show the corresponding estimated halo-halo skew-spectra in Figure~\ref{fig:clhhMCMC}, where one can see that the qualitative shapes of the spectra are recovered in the model within error bars. While we find that it is difficult to constrain the parameter $A_{\rm shot}$ due to the small amplitude of shot noise for our halo sample, we see that the data favour a sub-Poissonian $\ashot$, i.e. the shot noise is lower than the Poisson prediction. It is possible to model the shot noise differently. For example, we can vary two parameters $\alpha_1$ and $\alpha_2$ which measure the deviation of the $\bar{n}^{-1}$ and $\bar{n}^{-2}$ terms respectively \cite{MoradinezhadDizgah:2019xun,Schmittfull:2014tca}. Alternatively, one can restrict to a single parameter and include higher-order corrections of $\ashot$ \cite{Dai:2020adm} in the halo bispectrum. However, we see that these choices do not significantly affect the constraints on the biases. 

We find that $\bar{b}_1$ is the best constrained halo bias parameter since it enters as an overall amplitude in the halo $\times$ halo and halo $\times$ CMB lensing skew-spectra, and the percent error is $\sim 7.5\%$. Qualitatively, the correlations between $\bar{b}_1$, $\bar{b}_2$, and $\bar{b}_{\mathcal{G}_2}$ are as expected--- an increase in $\bar{b}_1$ must be compensated by a decrease in the others. Finally, we compare our posterior $1\sigma$ errors for the bias parameters with a Fisher forecast and find agreement at a 20\% level. Moreover, we observe 67\%, 36\% and 35\% tighter constraints for $\bar{b}_1$, $\bar{b}_2$ and $\bar{b}_{\G_2}$, respectively, when including the cross-correlation of the halo density field with CMB lensing. This is expected since our $hh\kappa$ and $\kappa\kappa h$ skew-spectra are sufficient to independently constrain all the bias parameters and $\ashot$. Note that we expect the comparison to change when we vary the cosmological parameters. However, our example already demonstrates how inclusion of the cross-correlation can improve constraints.

While we are able to obtain percent level errors on $\bar{b}_1$, we should keep in mind that there are a number of complications that can impact this result. For example, we assume Gaussian covariance and likelihood. We also consider a single-parameter scale-independent shot noise model. Furthermore, it is possible to improve our modelling of the skew-spectra, which is limited primarily by modelling challenges of the non-linear structure at $k\gtrsim0.4 h^{-1}$Mpc. Alternatively, access to data with a larger sky coverage would increase the number of modes and this should improve our results.

\begin{table}[h!]
    \centering
    \begin{tabular}{ccccc}
        \hline
        Parameter & $\bar{b}_1$ & $\bar{b}_2$ & $\bar{b}_{\G_2}$ & $A_{\rm shot}$ \\
        \hline
        Uniform Prior & $[1,\,3]$ & $[-2,\,12]$ & $[-5,\,5]$ & $[-1,\,1]$ \\
        \hline
        Halo + CMB Lensing & $1.203 \pm 0.091$ & $6.068 \pm 1.546$ & $1.111 \pm 0.562$ & $\geq0.581$ \\
        \hline
        Halo & $1.304 \pm 0.153$ & $4.502 \pm 2.104$ & $0.822 \pm 0.758$ & $\geq 0.006$\\
        \hline
    \end{tabular}
    \caption{Means and standard deviations from our MCMC likelihood analysis on the skew-spectra for $\kappa\kappa\kappa$, $ggg$, $\kappa$$gg$, and $g$$\kappa\kappa$, as well as for $hhh$ only. For $\ashot$, we show the $68\%$ lower bound instead. We use uniform priors on all the four parameters. We note that the $1\sigma$ errors for the bias parameters agree with the Fisher forecasted values (for the same configurations) at the 20\% level.}
    \label{tab:MCMCPrior}
\end{table}

%%%%%%%%%%%%%%%%%%%%%%%%%%%%%%%%%%%
\section{Conclusions}\label{sec: conclusion}

We show in this paper that we can efficiently extract the cosmological information encoded in the bispectra of the cross-correlation between CMB lensing and the galaxy/halo field. We do so by introducing the bispectrum-weighted skew-spectrum as a proxy statistic to compress the full cross-correlated bispectra in LSS $\times$ CMB lensing analysis. The problem thereafter simply becomes the estimation of the cross-correlated power spectra between a filtered quadratic field and a filtered density fields. 

This procedure utilizes the separable form of the angular bispectrum given in equation~\ref{eq:bispectrum} and the fact that convolution in harmonic space can be done by simple products in position space.

To show the power of the skew-spectrum, we demonstrate through a Fisher forecast that the information content of the bias parameters from the skew-spectrum is the same as that from the full bispectrum within $7\%$. We then move a step further to include the seven $\nu\Lambda$CDM cosmological parameters in our analysis and still find satisfactory agreement. Based on these results, we perform a forecast for DESI ELG $\times$ {\it Planck} lensing and find that the agreement is even better when we have more tomographic bins and the cross-correlation with CMB lensing. The differences between skew-spectra and full bispectra are less than $17\%$ for most of the parameters we consider in our analysis, reaching at most 12\% for all bias parameters.

We provide a fast pipeline to estimate the skew-spectrum from real data and, as a first step before application to real data, we apply our formalism to the {\tt MassiveNuS} simulation suite. We perform an MCMC likelihood analysis on the halo bias parameters and $A_{\rm shot}$ with the use of cross-correlated skew-spectra for the projected halo density field and the CMB lensing convergence field. Due to limitations in the available modes from the simulations we obtain modest constraints, especially on $\ashot$, a detection of which is limited by the amplitude of shot noise for our range of halo masses. Note that we assume a simple scenario in this analysis. For instance, we consider only a Gaussian likelihood and covariance. We should keep in mind that more rigorous studies on impacts from these assumptions are required before actual data analysis.

Realistic data will be affected by RSD and at small scales will be sensitive to loop corrections due to gravitational non-linearities and higher-order terms in the local bias expansion. A careful treatment of these effects is in place before applying our technique to real data. It will also be interesting to look into the kurt-spectra~\cite{Munshi:2021uwn} for higher-point correlation functions. We leave these for future work.

%%%%%%%%%%%%%%%%%%%%%%%%%%%%%%%%%%%
\acknowledgments
CD and SFC are partially supported by the Department of Energy (DOE) Grant No. DE-SC0020223.
We thank Tanveer Karim, Hayden Lee and Azadeh Moradinezhad Dizgah for useful discussions and insightful comments on the paper. We are also grateful to Antony Lewis for pointing out a typo in the previous version of this paper.
We acknowledge the use of {\tt CAMB}\footnote{\url{https://camb.info}}~\cite{Lewis:1999bs} and {\tt quicklens}~\cite{Planck:2015mym} for theoretical computations; {\tt LensTools}~\cite{Petri:2016zrq} and {\tt Halotools}~\cite{Hearin:2016uxs} for analysis of the simulation from {\tt MassiveNuS}~\cite{Liu:2017now}; {\tt emcee}~\cite{2013PASP..125..306F} for parameter estimation and {\tt GetDist}\footnote{\url{https://getdist.readthedocs.io/en/latest}}~\cite{Lewis:2019xzd} for generation of probability distribution plots.

\clearpage

\appendix

%%%%%%%%%%%%%%%%%%%%%%%%%%%%%%%%%%%
\section{Modelling of tracers}\label{A: tracer}

In this appendix we lay out the mapping that we use between matter overdensities and halo/galaxy overdensities, namely the model for bias and shot noise.

%%%%%%%%%%%%%%%%%%%%%%%%%%%%%%%%%%%%%%%%
\subsection{Bias}\label{A: bias}

In order to compute correlation functions of tracer fluctuations, it is necessary to invoke a bias model, i.e. a perturbative model that connects the tracer to the underlying matter fluctuations. We consider the basis of renormalized bias expansion in Eulerian space~\cite{Assassi:2015fma}, which only requires up to quadratic terms for the tree-level bispectrum, namely,
\begin{align}
    \delta_g &= b_1\delta_{cb}+\frac{b_2}{2}\delta_{cb}^2+b_{\G_2}\G_2[\Phi_{cb}]+\O(\Phi_{cb}^3)\,,
\end{align}
where $\G_2[\Phi_{cb}]=(\partial_i\partial_j\Phi_{cb})^2-(\partial^2\Phi_{cb})^2$ is the Galileon operator and $\Phi_{cb}$ is the gravitational potential due to the CDM and baryons alone~\cite{Villaescusa-Navarro:2013pva,Castorina:2013wga}. Assuming a local Lagrangian bias model~\cite{Desjacques:2016bnm} and using N-body simulations~\cite{Lazeyras:2015lgp}, it is possible to obtain fitting formulas for $b_2$ and $b_{\G_2}$ given prior knowledge of $b_1$. The relations are given by
\begin{align}
    b_1(z) &= \bar{b}_1\,p(z)\,,\\
    b_2(z) &= \bar{b}_2\left(0.412 - 2.143\,p(z) + 0.929\,p(z)^2 + 0.008\,p(z)^3\right)\,, \\
    b_{\G_2}(z) &= \bar{b}_{\G_2}\left(0.423 - 1.000\,p(z) + 0.310\,p(z)^2 +0.003\,p(z)^3 \right)\,,
\end{align}
where $\bar{b}_1$, $\bar{b}_2$ and $\bar{b}_{\G_2}$ are parameters for the overall amplitudes with fiducial values being equal to $1$ in our forecasts. Here $p(z)$ denotes the redshift evolution of the linear bias. 

%%%%%%%%%%%%%%%%%%%%%%%%%%%%%%%%%%%%%%%%%%%%%%
\subsection{Shot Noise}\label{A: shot noise}

The assumption that the halo catalogue is a result of a discrete Poisson sampling of the matter density field leads to the well-known shot noise in the halo power spectrum and bispectrum \cite{1980lssu.book.....P,jeong2010cosmology},
\begin{align}
    P_{\text{Poisson}} &= \frac{1}{\Bar{n}_h} \\
    B_{\text{Poisson}}(k_1,k_2,k_3) &= \frac{1}{\Bar{n}_h}\left(P_{hh}(k_1)+P_{hh}(k_2)+P_{hh}(k_3)\right)+\frac{1}{\Bar{n}_h^2}\,,
\end{align}
where $\Bar{n}_h$ is the average number density of halos (or galaxies) in a given survey volume. However, the Poisson prediction is not exact due to halo clustering and exclusion effects, as shown in \cite{Baldauf:2013hka}. These effects either raise or lower the Poisson shot noise at large scales (referred to as super- or sub-Poissonian) and can be modelled phenomenologically as in \cite{Schmittfull:2014tca,Dai:2020adm,Gil-Marin:2014sta}. In these studies, a parameter $\ashot$ was introduced to measure deviations from the Poisson prediction. The model we use is as follows,
\begin{align}
    P_{\text{shot}}(k) &= P_{\text{Poisson}}(1-\ashot)\,, \\
    B_{\text{shot}}(k) &= B_{\text{Poisson}}(1-\ashot)\,,
\end{align}
with $\abs{\ashot}\leq1$.
Here, $\ashot$ simply measures whether the true shot noise is higher or lower than the Poisson prediction, and remains scale-independent. 

%%%%%%%%%%%%%%%%%%%%%%%%%%%%%%%%%%%%%%%%%%%%%
\section{Survey Specifications}\label{A: survey specs}

In this appendix we detail the experimental specifications for our forecasts in Section~\ref{sec: forecast}. This includes DESI and {\it Planck}.

%%%%%%%%%%%%%%%%%%%%%%%%%%%%%%%%%%%%%%%%%%%%%%%%
\subsection{DESI}

We consider the emission line galaxies (ELG) in our study~\cite{DESI:2016fyo}. The sky coverage for DESI is 14000 $\text{deg}^2$, which corresponds to $f_{\rm sky}=0.34$. Table~\ref{tab:DESI} shows the number density of DESI in each redshift \cite{DESI:2016fyo}. The galaxy bias is modelled as 
\begin{align}
    b_1(k,z) = 1.46\times0.84\,\bar{b}_1 D(k,z)^{-1}\,,
\end{align}
where the factor 1.46 is chosen such that the linear bias agrees with simulations at $z=0$ (for halo masses being $3\times10^{13}\,h^{-1}M_{\odot}$~\cite{Lazeyras:2015lgp}) and $\bar{b}_1$ is the parameter that parametrizes the overall amplitude used in our forecast. 

We split the redshift range into four tomographic bins (equally spaced in redshift). If we consider $k_{\rm max}=0.1$ $h{\rm Mpc}^{-1}$, then the $\ell_{\rm max}$ for the first and last bins are 189 and 305, respectively. But for simplicity, we choose a fixed $\ell_{\rm max}=305$ for all four bins.
\begin{table}[!h]
    \centering
    \begin{tabular}{cccccccccccc}
    \toprule
    $z$ & 0.65 & 0.75 & 0.85 & 0.95 & 1.05 & 1.15 & 1.25 & 1.35 & 1.45 & 1.55 & 1.65 \\
    \midrule
    $\d N/\d z\,\d\text{deg}^{2}$ & 309 & 2269 & 1923 & 2094 & 1441 & 1353 & 1337 & 563 & 466 & 329 & 126\\
    \bottomrule
    \end{tabular} 
    \caption{Experimental specification for DESI ELG \cite{DESI:2016fyo}.}
    \label{tab:DESI}
\end{table}

%%%%%%%%%%%%%%%%%%%%%%%%%%%%%%%%%%%%%%%%%%%%%%%%
\subsection{Planck}

We model the experimental configuration of {\it Planck} as one effective frequency to approximate the complicated {\it Planck} likelihood. We take $\Delta_T=43\,\mu {\rm K}'$, $\Delta_P=81\,\mu {\rm K}'$ and $\theta_{\rm FWHM}=5'$ as our noise model. For the multipole range, we consider $\ell\in[2,\,2500]$ for both temperature and polarization maps. With these values, we then use the public code {\tt quicklens}\footnote{\url{https://github.com/dhanson/quicklens}}~\cite{Planck:2015mym} to compute the lensing reconstruction noise from the minimum variance quadratic estimator. In order to perform the cross-correlation appropriately, we assume the same $f_{\rm sky}$ and multipole range as DESI ELG for {\it Planck} lensing convergence maps.

%%%%%%%%%%%%%%%%%%%%%%%%%%%%%%%%%%%%%%%
\section{Covariance in Flat Sky}\label{A:Covariance}

\begin{figure}[h!]
    \centering
    \includegraphics[scale=1.4]{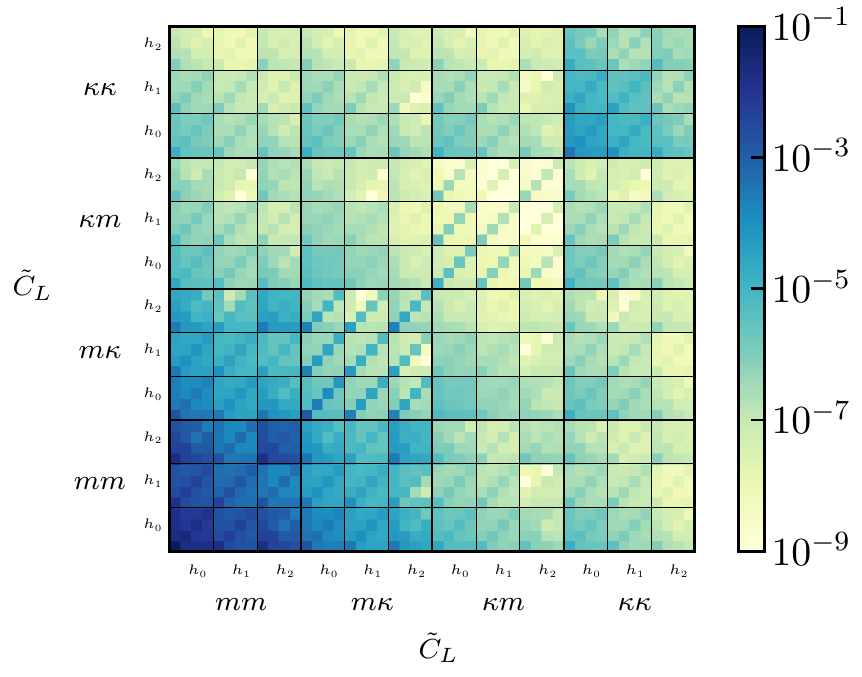}
    \caption{Theoretical evaluation of the flat-sky covariance matrix used in our MCMC analysis, as discussed in Section~\ref{sec:flatsky}. The $h_i$s denotes the kernel from the $i$-th Legendre polynomial. The elements in each block matrix are ordered by increasing value of $L$. In our MCMC analysis, we rescale each submatrix with the corresponding power of the linear bias $b_1$ and add contributions from shot noise (scaled by $\ashot$).}
    \label{fig:covarianceFlatTheoretical}
\end{figure}

As a validation of our methodology, besides the shape of skew-spectra themselves, we show in this appendix a comparison between the covariance matrix using the theoretical prediction and the simulations. Before presenting the comparison, let us first take a look at Figure~\ref{fig:covarianceFlatTheoretical}, which shows the theoretical covariance matrix using the flat-sky approximation. Different from the full-sky covariance shown in Section~\ref{sec:Leg Pol Dec}, the amplitude of off-diagonal terms is comparable to the diagonal pieces. 

\begin{figure}[h!]
    \centering
    \includegraphics[scale=1.]{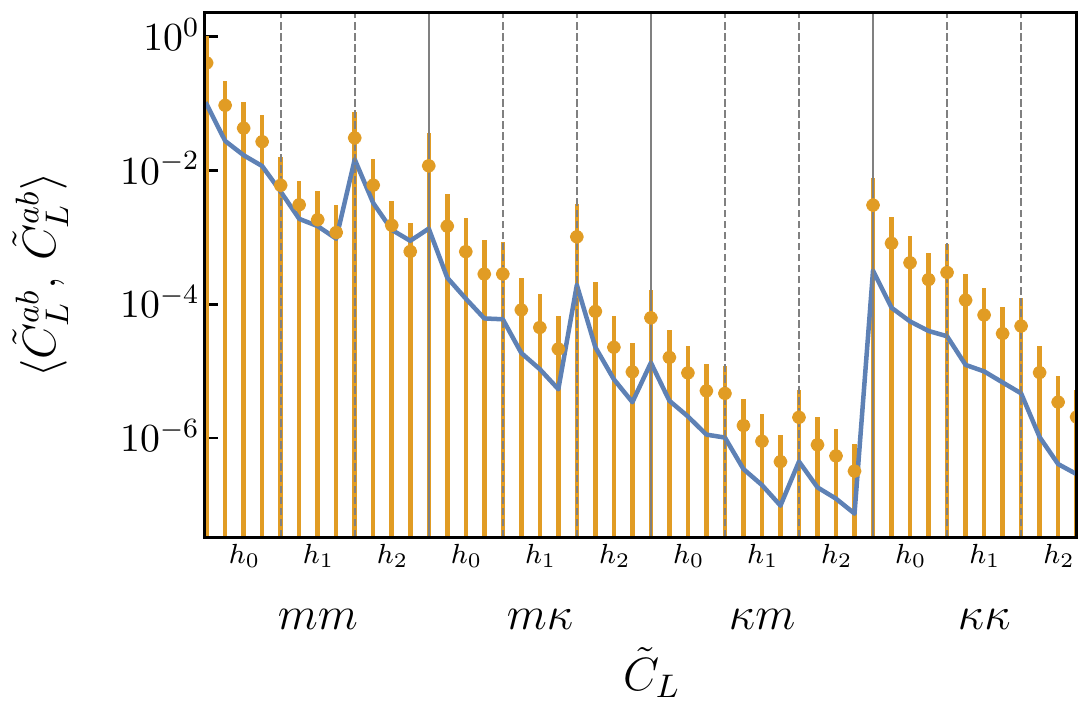}
    \caption{Comparison between the diagonal terms of the simulated and the theoretical covariance matrices in flat sky, where $\{a,\,b\}\in\{m,\,\kappa\}$. The blue line shows the theoretical computation using equations in Section~\ref{sec:flatsky}. The orange dots show the average value of the estimated covariance from the simulations and the error bars represent their variance. The $h_i$s denotes the kernel from the $i$-th Legendre polynomial. The bins for the kernels correspond to increasing values of $L$.}
    \label{fig:covarianceFlatSimulation}
\end{figure}

We analyze the same simulation set described in Section~\ref{sec: sims} and compute the estimated covariance by averaging over all our realizations of lensing convergence maps from one simulation box. We show our comparison for the diagonal terms in Figure~\ref{fig:covarianceFlatSimulation}. We see that we can reproduce the trends of the simulation with only the Gaussian (disconnected) part of the theoretical covariance. We also observe that for the cases including $\kappa\kappa$, $\kappa m$ and $m\kappa$, the simulation has a larger disagreement with our theoretical modelling. This is expected since the fitting function we use to model the matter bispectrum is insufficient at the small scales probed by the lensing convergence. 


\bibliographystyle{utphys}
\bibliography{refs.bib}

\providecommand{\href}[2]{#2}\begingroup\raggedright\begin{thebibliography}{10}

\bibitem{DESI:2016fyo}
{\bfseries DESI} Collaboration, A.~Aghamousa {\em et~al.}, ``{The DESI
  Experiment Part I: Science,Targeting, and Survey Design},''
  \href{http://arxiv.org/abs/1611.00036}{{\ttfamily arXiv:1611.00036
  [astro-ph.IM]}}.

\bibitem{Abell:2009aa}
{\bfseries LSST Science, LSST Project} Collaboration, P.~A. Abell {\em et~al.},
  ``{LSST Science Book, Version 2.0},''
  \href{http://arxiv.org/abs/0912.0201}{{\ttfamily arXiv:0912.0201
  [astro-ph.IM]}}.

\bibitem{Planck:2018vyg}
{\bfseries Planck} Collaboration, N.~Aghanim {\em et~al.}, ``{Planck 2018
  results. VI. Cosmological parameters},''
  \href{http://dx.doi.org/10.1051/0004-6361/201833910}{{\em Astron. Astrophys.}
  {\bfseries 641} (2020) A6}, \href{http://arxiv.org/abs/1807.06209}{{\ttfamily
  arXiv:1807.06209 [astro-ph.CO]}}. [Erratum: Astron.Astrophys. 652, C4
  (2021)].

\bibitem{CMB-S4:2016ple}
{\bfseries CMB-S4} Collaboration, K.~N. Abazajian {\em et~al.}, ``{CMB-S4
  Science Book, First Edition},''
  \href{http://arxiv.org/abs/1610.02743}{{\ttfamily arXiv:1610.02743
  [astro-ph.CO]}}.

\bibitem{SimonsObservatory:2018koc}
{\bfseries Simons Observatory} Collaboration, P.~Ade {\em et~al.}, ``{The
  Simons Observatory: Science goals and forecasts},''
  \href{http://dx.doi.org/10.1088/1475-7516/2019/02/056}{{\em JCAP} {\bfseries
  02} (2019) 056}, \href{http://arxiv.org/abs/1808.07445}{{\ttfamily
  arXiv:1808.07445 [astro-ph.CO]}}.

\bibitem{Yu:2021vce}
B.~Yu, S.~Ferraro, Z.~R. Knight, L.~Knox, and B.~D. Sherwin, ``{The Physical
  Origin of Dark Energy Constraints from Rubin Observatory and CMB-S4 Lensing
  Tomography},'' \href{http://arxiv.org/abs/2108.02801}{{\ttfamily
  arXiv:2108.02801 [astro-ph.CO]}}.

\bibitem{Pullen:2015vtb}
A.~R. Pullen, S.~Alam, S.~He, and S.~Ho, ``{Constraining Gravity at the Largest
  Scales through CMB Lensing and Galaxy Velocities},''
  \href{http://dx.doi.org/10.1093/mnras/stw1249}{{\em Mon. Not. Roy. Astron.
  Soc.} {\bfseries 460} no.~4, (2016) 4098--4108},
  \href{http://arxiv.org/abs/1511.04457}{{\ttfamily arXiv:1511.04457
  [astro-ph.CO]}}.

\bibitem{Seljak:2008xr}
U.~Seljak, ``{Extracting primordial non-gaussianity without cosmic variance},''
  \href{http://dx.doi.org/10.1103/PhysRevLett.102.021302}{{\em Phys. Rev.
  Lett.} {\bfseries 102} (2009) 021302},
  \href{http://arxiv.org/abs/0807.1770}{{\ttfamily arXiv:0807.1770
  [astro-ph]}}.

\bibitem{Schmittfull:2017ffw}
M.~Schmittfull and U.~Seljak, ``{Parameter constraints from cross-correlation
  of CMB lensing with galaxy clustering},''
  \href{http://dx.doi.org/10.1103/PhysRevD.97.123540}{{\em Phys. Rev. D}
  {\bfseries 97} no.~12, (2018) 123540},
  \href{http://arxiv.org/abs/1710.09465}{{\ttfamily arXiv:1710.09465
  [astro-ph.CO]}}.

\bibitem{Yu:2018tem}
B.~Yu, R.~Z. Knight, B.~D. Sherwin, S.~Ferraro, L.~Knox, and M.~Schmittfull,
  ``{Towards Neutrino Mass from Cosmology without Optical Depth Information},''
  \href{http://arxiv.org/abs/1809.02120}{{\ttfamily arXiv:1809.02120
  [astro-ph.CO]}}.

\bibitem{DES:2018miq}
{\bfseries DES, SPT} Collaboration, Y.~Omori {\em et~al.}, ``{Dark Energy
  Survey Year 1 Results: Tomographic cross-correlations between Dark Energy
  Survey galaxies and CMB lensing from South Pole Telescope+Planck},''
  \href{http://dx.doi.org/10.1103/PhysRevD.100.043501}{{\em Phys. Rev. D}
  {\bfseries 100} no.~4, (2019) 043501},
  \href{http://arxiv.org/abs/1810.02342}{{\ttfamily arXiv:1810.02342
  [astro-ph.CO]}}.

\bibitem{Zhang:2021eqp}
{\bfseries LSST Dark Energy Science} Collaboration, Z.~Zhang, C.~Chang,
  P.~Larsen, L.~F. Secco, and J.~Zuntz, ``{Transitioning from Stage-III to
  Stage-IV: Cosmology from galaxy$\times$CMB lensing and shear$\times$CMB
  lensing},'' \href{http://arxiv.org/abs/2111.04917}{{\ttfamily
  arXiv:2111.04917 [astro-ph.CO]}}.

\bibitem{Sgier:2021bzf}
R.~Sgier, C.~Lorenz, A.~Refregier, J.~Fluri, D.~Z\"urcher, and F.~Tarsitano,
  ``{Combined $13\times2$-point analysis of the Cosmic Microwave Background and
  Large-Scale Structure: implications for the $S_8$-tension and neutrino mass
  constraints},'' \href{http://arxiv.org/abs/2110.03815}{{\ttfamily
  arXiv:2110.03815 [astro-ph.CO]}}.

\bibitem{Sailer:2021yzm}
N.~Sailer, E.~Castorina, S.~Ferraro, and M.~White, ``{Cosmology at high
  redshift \textemdash{} a probe of fundamental physics},''
  \href{http://dx.doi.org/10.1088/1475-7516/2021/12/049}{{\em JCAP} {\bfseries
  12} no.~12, (2021) 049}, \href{http://arxiv.org/abs/2106.09713}{{\ttfamily
  arXiv:2106.09713 [astro-ph.CO]}}.

\bibitem{White:2021yvw}
M.~White {\em et~al.}, ``{Cosmological constraints from the tomographic
  cross-correlation of DESI Luminous Red Galaxies and Planck CMB lensing},''
  \href{http://arxiv.org/abs/2111.09898}{{\ttfamily arXiv:2111.09898
  [astro-ph.CO]}}.

\bibitem{Cooray:2001ps}
A.~Cooray, ``{Squared temperature-temperature power spectrum as a probe of the
  CMB bispectrum},'' \href{http://dx.doi.org/10.1103/PhysRevD.64.043516}{{\em
  Phys. Rev. D} {\bfseries 64} (2001) 043516},
  \href{http://arxiv.org/abs/astro-ph/0105415}{{\ttfamily
  arXiv:astro-ph/0105415}}.

\bibitem{Babich:2004yc}
D.~Babich and M.~Zaldarriaga, ``{Primordial bispectrum information from CMB
  polarization},'' \href{http://dx.doi.org/10.1103/PhysRevD.70.083005}{{\em
  Phys. Rev. D} {\bfseries 70} (2004) 083005},
  \href{http://arxiv.org/abs/astro-ph/0408455}{{\ttfamily
  arXiv:astro-ph/0408455}}.

\bibitem{Komatsu:2003iq}
E.~Komatsu, D.~N. Spergel, and B.~D. Wandelt, ``{Measuring primordial
  non-Gaussianity in the cosmic microwave background},''
  \href{http://dx.doi.org/10.1086/491724}{{\em Astrophys. J.} {\bfseries 634}
  (2005) 14--19}, \href{http://arxiv.org/abs/astro-ph/0305189}{{\ttfamily
  arXiv:astro-ph/0305189}}.

\bibitem{Schmittfull:2014tca}
M.~Schmittfull, T.~Baldauf, and U.~Seljak, ``{Near optimal bispectrum
  estimators for large-scale structure},''
  \href{http://dx.doi.org/10.1103/PhysRevD.91.043530}{{\em Phys. Rev. D}
  {\bfseries 91} no.~4, (2015) 043530},
  \href{http://arxiv.org/abs/1411.6595}{{\ttfamily arXiv:1411.6595
  [astro-ph.CO]}}.

\bibitem{MoradinezhadDizgah:2019xun}
A.~Moradinezhad~Dizgah, H.~Lee, M.~Schmittfull, and C.~Dvorkin, ``{Capturing
  non-Gaussianity of the large-scale structure with weighted skew-spectra},''
  \href{http://dx.doi.org/10.1088/1475-7516/2020/04/011}{{\em JCAP} {\bfseries
  04} (2020) 011}, \href{http://arxiv.org/abs/1911.05763}{{\ttfamily
  arXiv:1911.05763 [astro-ph.CO]}}.

\bibitem{Schmittfull:2020hoi}
M.~Schmittfull and A.~Moradinezhad~Dizgah, ``{Galaxy skew-spectra in
  redshift-space},''
  \href{http://dx.doi.org/10.1088/1475-7516/2021/03/020}{{\em JCAP} {\bfseries
  03} (2021) 020}, \href{http://arxiv.org/abs/2010.14267}{{\ttfamily
  arXiv:2010.14267 [astro-ph.CO]}}.

\bibitem{Munshi:2020ofi}
D.~Munshi, T.~Namikawa, T.~D. Kitching, J.~D. McEwen, and F.~R. Bouchet,
  ``{Weak Lensing Skew-Spectrum},''
  \href{http://dx.doi.org/10.1093/mnras/staa2769}{{\em Mon. Not. Roy. Astron.
  Soc.} {\bfseries 498} no.~4, (2020) 6057--6068},
  \href{http://arxiv.org/abs/2006.12832}{{\ttfamily arXiv:2006.12832
  [astro-ph.CO]}}.

\bibitem{Olive:2016}
K.~Olive, ``Review of particle physics,''
  \href{http://dx.doi.org/10.1088/1674-1137/40/10/100001}{{\em Chinese Physics
  C} {\bfseries 40} no.~10, (Oct, 2016) 100001}.
  \url{https://doi.org/10.1088/1674-1137/40/10/100001}.

\bibitem{Dvorkin:2019jgs}
C.~Dvorkin {\em et~al.}, ``{Neutrino Mass from Cosmology: Probing Physics
  Beyond the Standard Model},''
  \href{http://arxiv.org/abs/1903.03689}{{\ttfamily arXiv:1903.03689
  [astro-ph.CO]}}.

\bibitem{Chen:2021vba}
S.-F. Chen, H.~Lee, and C.~Dvorkin, ``{Precise and accurate cosmology with
  CMB\texttimes{}LSS power spectra and bispectra},''
  \href{http://dx.doi.org/10.1088/1475-7516/2021/05/030}{{\em JCAP} {\bfseries
  05} (2021) 030}, \href{http://arxiv.org/abs/2103.01229}{{\ttfamily
  arXiv:2103.01229 [astro-ph.CO]}}.

\bibitem{Liu:2017now}
J.~Liu, S.~Bird, J.~M.~Z. Matilla, J.~C. Hill, Z.~Haiman, M.~S. Madhavacheril,
  A.~Petri, and D.~N. Spergel, ``{MassiveNuS: Cosmological Massive Neutrino
  Simulations},'' \href{http://dx.doi.org/10.1088/1475-7516/2018/03/049}{{\em
  JCAP} {\bfseries 03} (2018) 049},
  \href{http://arxiv.org/abs/1711.10524}{{\ttfamily arXiv:1711.10524
  [astro-ph.CO]}}.

\bibitem{Dai:2020adm}
J.-P. Dai, L.~Verde, and J.-Q. Xia, ``{What Can We Learn by Combining the Skew
  Spectrum and the Power Spectrum?},''
  \href{http://dx.doi.org/10.1088/1475-7516/2020/08/007}{{\em JCAP} {\bfseries
  08} (2020) 007}, \href{http://arxiv.org/abs/2002.09904}{{\ttfamily
  arXiv:2002.09904 [astro-ph.CO]}}.

\bibitem{1980lssu.book.....P}
P.~J.~E. {Peebles}, {\em {The large-scale structure of the universe}}.
\newblock 1980.

\bibitem{Lewis:2006fu}
A.~Lewis and A.~Challinor, ``{Weak gravitational lensing of the CMB},''
  \href{http://dx.doi.org/10.1016/j.physrep.2006.03.002}{{\em Phys. Rept.}
  {\bfseries 429} (2006) 1--65},
  \href{http://arxiv.org/abs/astro-ph/0601594}{{\ttfamily
  arXiv:astro-ph/0601594}}.

\bibitem{Bartelmann:1999yn}
M.~Bartelmann and P.~Schneider, ``{Weak gravitational lensing},''
  \href{http://dx.doi.org/10.1016/S0370-1573(00)00082-X}{{\em Phys. Rept.}
  {\bfseries 340} (2001) 291--472},
  \href{http://arxiv.org/abs/astro-ph/9912508}{{\ttfamily
  arXiv:astro-ph/9912508}}.

\bibitem{Liu:2013yna}
J.~Liu, Z.~Haiman, L.~Hui, J.~M. Kratochvil, and M.~May, ``{The Impact of
  Magnification and Size Bias on Weak Lensing Power Spectrum and Peak
  Statistics},'' \href{http://dx.doi.org/10.1103/PhysRevD.89.023515}{{\em Phys.
  Rev. D} {\bfseries 89} no.~2, (2014) 023515},
  \href{http://arxiv.org/abs/1310.7517}{{\ttfamily arXiv:1310.7517
  [astro-ph.CO]}}.

\bibitem{Hui:2007cu}
L.~Hui, E.~Gaztanaga, and M.~LoVerde, ``{Anisotropic Magnification Distortion
  of the 3D Galaxy Correlation. 1. Real Space},''
  \href{http://dx.doi.org/10.1103/PhysRevD.76.103502}{{\em Phys. Rev. D}
  {\bfseries 76} (2007) 103502},
  \href{http://arxiv.org/abs/0706.1071}{{\ttfamily arXiv:0706.1071
  [astro-ph]}}.

\bibitem{Lee:2020ebj}
H.~Lee and C.~Dvorkin, ``{Cosmological Angular Trispectra and Non-Gaussian
  Covariance},'' \href{http://dx.doi.org/10.1088/1475-7516/2020/05/044}{{\em
  JCAP} {\bfseries 05} (2020) 044},
  \href{http://arxiv.org/abs/2001.00584}{{\ttfamily arXiv:2001.00584
  [astro-ph.CO]}}.

\bibitem{Limber:1954zz}
D.~N. Limber, ``{The Analysis of Counts of the Extragalactic Nebulae in Terms
  of a Fluctuating Density Field. II},''
  \href{http://dx.doi.org/10.1086/145870}{{\em Astrophys. J.} {\bfseries 119}
  (1954) 655}.

\bibitem{Assassi:2017lea}
V.~Assassi, M.~Simonovi\'c, and M.~Zaldarriaga, ``{Efficient evaluation of
  angular power spectra and bispectra},''
  \href{http://dx.doi.org/10.1088/1475-7516/2017/11/054}{{\em JCAP} {\bfseries
  11} (2017) 054}, \href{http://arxiv.org/abs/1705.05022}{{\ttfamily
  arXiv:1705.05022 [astro-ph.CO]}}.

\bibitem{Hamilton:1999uv}
A.~J.~S. Hamilton, ``{Uncorrelated modes of the nonlinear power spectrum},''
  \href{http://dx.doi.org/10.1046/j.1365-8711.2000.03071.x}{{\em Mon. Not. Roy.
  Astron. Soc.} {\bfseries 312} (2000) 257--284},
  \href{http://arxiv.org/abs/astro-ph/9905191}{{\ttfamily
  arXiv:astro-ph/9905191}}.

\bibitem{Senatore:2017hyk}
L.~Senatore and M.~Zaldarriaga, ``{The Effective Field Theory of Large-Scale
  Structure in the presence of Massive Neutrinos},''
  \href{http://arxiv.org/abs/1707.04698}{{\ttfamily arXiv:1707.04698
  [astro-ph.CO]}}.

\bibitem{Hu:1997vi}
W.~Hu and D.~J. Eisenstein, ``{Small scale perturbations in a general MDM
  cosmology},'' \href{http://dx.doi.org/10.1086/305585}{{\em Astrophys. J.}
  {\bfseries 498} (1998) 497},
  \href{http://arxiv.org/abs/astro-ph/9710216}{{\ttfamily
  arXiv:astro-ph/9710216}}.

\bibitem{Takada:2005si}
M.~Takada, E.~Komatsu, and T.~Futamase, ``{Cosmology with high-redshift galaxy
  survey: neutrino mass and inflation},''
  \href{http://dx.doi.org/10.1103/PhysRevD.73.083520}{{\em Phys. Rev. D}
  {\bfseries 73} (2006) 083520},
  \href{http://arxiv.org/abs/astro-ph/0512374}{{\ttfamily
  arXiv:astro-ph/0512374}}.

\bibitem{Lesgourgues:2006nd}
J.~Lesgourgues and S.~Pastor, ``{Massive neutrinos and cosmology},''
  \href{http://dx.doi.org/10.1016/j.physrep.2006.04.001}{{\em Phys. Rept.}
  {\bfseries 429} (2006) 307--379},
  \href{http://arxiv.org/abs/astro-ph/0603494}{{\ttfamily
  arXiv:astro-ph/0603494}}.

\bibitem{Berlind:2001xk}
A.~A. Berlind and D.~H. Weinberg, ``{The Halo occupation distribution: Towards
  an empirical determination of the relation between galaxies and mass},''
  \href{http://dx.doi.org/10.1086/341469}{{\em Astrophys. J.} {\bfseries 575}
  (2002) 587--616}, \href{http://arxiv.org/abs/astro-ph/0109001}{{\ttfamily
  arXiv:astro-ph/0109001}}.

\bibitem{Desjacques:2016bnm}
V.~Desjacques, D.~Jeong, and F.~Schmidt, ``{Large-Scale Galaxy Bias},''
  \href{http://dx.doi.org/10.1016/j.physrep.2017.12.002}{{\em Phys. Rept.}
  {\bfseries 733} (2018) 1--193},
  \href{http://arxiv.org/abs/1611.09787}{{\ttfamily arXiv:1611.09787
  [astro-ph.CO]}}.

\bibitem{Tegmark:1997rp}
M.~Tegmark, ``{Measuring cosmological parameters with galaxy surveys},''
  \href{http://dx.doi.org/10.1103/PhysRevLett.79.3806}{{\em Phys. Rev. Lett.}
  {\bfseries 79} (1997) 3806--3809},
  \href{http://arxiv.org/abs/astro-ph/9706198}{{\ttfamily
  arXiv:astro-ph/9706198}}.

\bibitem{Yadav:2007rk}
A.~P.~S. Yadav, E.~Komatsu, and B.~D. Wandelt, ``{Fast Estimator of Primordial
  Non-Gaussianity from Temperature and Polarization Anisotropies in the Cosmic
  Microwave Background},'' \href{http://dx.doi.org/10.1086/519071}{{\em
  Astrophys. J.} {\bfseries 664} (2007) 680--686},
  \href{http://arxiv.org/abs/astro-ph/0701921}{{\ttfamily
  arXiv:astro-ph/0701921}}.

\bibitem{Baldauf:2013hka}
T.~Baldauf, U.~Seljak, R.~E. Smith, N.~Hamaus, and V.~Desjacques, ``{Halo
  stochasticity from exclusion and nonlinear clustering},''
  \href{http://dx.doi.org/10.1103/PhysRevD.88.083507}{{\em Phys. Rev. D}
  {\bfseries 88} no.~8, (2013) 083507},
  \href{http://arxiv.org/abs/1305.2917}{{\ttfamily arXiv:1305.2917
  [astro-ph.CO]}}.

\bibitem{Gil-Marin:2014sta}
H.~Gil-Mar\'\i{}n, J.~Nore\~na, L.~Verde, W.~J. Percival, C.~Wagner, M.~Manera,
  and D.~P. Schneider, ``{The power spectrum and bispectrum of SDSS DR11 BOSS
  galaxies \textendash{} I. Bias and gravity},''
  \href{http://dx.doi.org/10.1093/mnras/stv961}{{\em Mon. Not. Roy. Astron.
  Soc.} {\bfseries 451} no.~1, (2015) 539--580},
  \href{http://arxiv.org/abs/1407.5668}{{\ttfamily arXiv:1407.5668
  [astro-ph.CO]}}.

\bibitem{Smith:2002dz}
{\bfseries VIRGO Consortium} Collaboration, R.~E. Smith, J.~A. Peacock,
  A.~Jenkins, S.~D.~M. White, C.~S. Frenk, F.~R. Pearce, P.~A. Thomas,
  G.~Efstathiou, and H.~M.~P. Couchmann, ``{Stable clustering, the halo model
  and nonlinear cosmological power spectra},''
  \href{http://dx.doi.org/10.1046/j.1365-8711.2003.06503.x}{{\em Mon. Not. Roy.
  Astron. Soc.} {\bfseries 341} (2003) 1311},
  \href{http://arxiv.org/abs/astro-ph/0207664}{{\ttfamily
  arXiv:astro-ph/0207664}}.

\bibitem{Mead:2020vgs}
A.~Mead, S.~Brieden, T.~Tr\"oster, and C.~Heymans, ``{HMcode-2020: Improved
  modelling of non-linear cosmological power spectra with baryonic feedback},''
  \href{http://arxiv.org/abs/2009.01858}{{\ttfamily arXiv:2009.01858
  [astro-ph.CO]}}.

\bibitem{Gil-Marin:2011jtv}
H.~Gil-Marin, C.~Wagner, F.~Fragkoudi, R.~Jimenez, and L.~Verde, ``{An improved
  fitting formula for the dark matter bispectrum},''
  \href{http://dx.doi.org/10.1088/1475-7516/2012/02/047}{{\em JCAP} {\bfseries
  02} (2012) 047}, \href{http://arxiv.org/abs/1111.4477}{{\ttfamily
  arXiv:1111.4477 [astro-ph.CO]}}.

\bibitem{Takahashi:2019hth}
R.~Takahashi, T.~Nishimichi, T.~Namikawa, A.~Taruya, I.~Kayo, K.~Osato,
  Y.~Kobayashi, and M.~Shirasaki, ``{Fitting the nonlinear matter bispectrum by
  the Halofit approach},''
  \href{http://dx.doi.org/10.3847/1538-4357/ab908d}{{\em Astrophys. J.}
  {\bfseries 895} no.~2, (2020) 113},
  \href{http://arxiv.org/abs/1911.07886}{{\ttfamily arXiv:1911.07886
  [astro-ph.CO]}}.

\bibitem{Pratten:2016dsm}
G.~Pratten and A.~Lewis, ``{Impact of post-Born lensing on the CMB},''
  \href{http://dx.doi.org/10.1088/1475-7516/2016/08/047}{{\em JCAP} {\bfseries
  08} (2016) 047}, \href{http://arxiv.org/abs/1605.05662}{{\ttfamily
  arXiv:1605.05662 [astro-ph.CO]}}.

\bibitem{Fabbian:2019tik}
G.~Fabbian, A.~Lewis, and D.~Beck, ``{CMB lensing reconstruction biases in
  cross-correlation with large-scale structure probes},''
  \href{http://dx.doi.org/10.1088/1475-7516/2019/10/057}{{\em JCAP} {\bfseries
  10} (2019) 057}, \href{http://arxiv.org/abs/1906.08760}{{\ttfamily
  arXiv:1906.08760 [astro-ph.CO]}}.

\bibitem{DiDio:2018unb}
E.~Di~Dio, R.~Durrer, R.~Maartens, F.~Montanari, and O.~Umeh, ``{The Full-Sky
  Angular Bispectrum in Redshift Space},''
  \href{http://dx.doi.org/10.1088/1475-7516/2019/04/053}{{\em JCAP} {\bfseries
  04} (2019) 053}, \href{http://arxiv.org/abs/1812.09297}{{\ttfamily
  arXiv:1812.09297 [astro-ph.CO]}}.

\bibitem{Hearin:2016uxs}
A.~P. Hearin {\em et~al.}, ``{Forward Modeling of Large-Scale Structure: An
  open-source approach with Halotools},''
  \href{http://dx.doi.org/10.5281/zenodo.835895}{{\em Astron. J.} {\bfseries
  154} no.~5, (2017) 190}, \href{http://arxiv.org/abs/1606.04106}{{\ttfamily
  arXiv:1606.04106 [astro-ph.IM]}}.

\bibitem{Petri:2016zrq}
A.~Petri, ``{Mocking the Weak Lensing universe: the LensTools python computing
  package},'' \href{http://dx.doi.org/10.1016/j.ascom.2016.06.001}{{\em Astron.
  Comput.} {\bfseries 17} (2016) 73--79},
  \href{http://arxiv.org/abs/1606.01903}{{\ttfamily arXiv:1606.01903
  [astro-ph.CO]}}.

\bibitem{Petri:2016wlu}
A.~Petri, Z.~Haiman, and M.~May, ``{Sample variance in weak lensing: how many
  simulations are required?},''
  \href{http://dx.doi.org/10.1103/PhysRevD.93.063524}{{\em Phys. Rev. D}
  {\bfseries 93} no.~6, (2016) 063524},
  \href{http://arxiv.org/abs/1601.06792}{{\ttfamily arXiv:1601.06792
  [astro-ph.CO]}}.

\bibitem{Coulton:2018ebd}
W.~R. Coulton, J.~Liu, M.~S. Madhavacheril, V.~B\"ohm, and D.~N. Spergel,
  ``{Constraining Neutrino Mass with the Tomographic Weak Lensing
  Bispectrum},'' \href{http://dx.doi.org/10.1088/1475-7516/2019/05/043}{{\em
  JCAP} {\bfseries 05} (2019) 043},
  \href{http://arxiv.org/abs/1810.02374}{{\ttfamily arXiv:1810.02374
  [astro-ph.CO]}}.

\bibitem{Bucher:2009nm}
M.~Bucher, B.~Van~Tent, and C.~S. Carvalho, ``{Detecting Bispectral Acoustic
  Oscillations from Inflation Using a New Flexible Estimator},''
  \href{http://dx.doi.org/10.1111/j.1365-2966.2010.17089.x}{{\em Mon. Not. Roy.
  Astron. Soc.} {\bfseries 407} (2010) 2193},
  \href{http://arxiv.org/abs/0911.1642}{{\ttfamily arXiv:0911.1642
  [astro-ph.CO]}}.

\bibitem{Bucher:2015ura}
M.~Bucher, B.~Racine, and B.~van Tent, ``{The binned bispectrum estimator:
  template-based and non-parametric CMB non-Gaussianity searches},''
  \href{http://dx.doi.org/10.1088/1475-7516/2016/05/055}{{\em JCAP} {\bfseries
  05} (2016) 055}, \href{http://arxiv.org/abs/1509.08107}{{\ttfamily
  arXiv:1509.08107 [astro-ph.CO]}}.

\bibitem{Hu:2000ee}
W.~Hu, ``{Weak lensing of the CMB: A harmonic approach},''
  \href{http://dx.doi.org/10.1103/PhysRevD.62.043007}{{\em Phys. Rev. D}
  {\bfseries 62} (2000) 043007},
  \href{http://arxiv.org/abs/astro-ph/0001303}{{\ttfamily
  arXiv:astro-ph/0001303}}.

\bibitem{2013PASP..125..306F}
D.~{Foreman-Mackey}, D.~W. {Hogg}, D.~{Lang}, and J.~{Goodman}, ``{emcee: The
  MCMC Hammer},'' \href{http://arxiv.org/abs/1202.3665}{{\ttfamily
  arXiv:1202.3665 [astro-ph.IM]}}.

\bibitem{Munshi:2021uwn}
D.~Munshi, H.~Lee, C.~Dvorkin, and J.~D. McEwen, ``{Weak Lensing Trispectrum
  and Kurt-Spectra},'' \href{http://arxiv.org/abs/2112.05155}{{\ttfamily
  arXiv:2112.05155 [astro-ph.CO]}}.

\bibitem{Lewis:1999bs}
A.~Lewis, A.~Challinor, and A.~Lasenby, ``{Efficient computation of CMB
  anisotropies in closed FRW models},''
  \href{http://dx.doi.org/10.1086/309179}{{\em Astrophys. J.} {\bfseries 538}
  (2000) 473--476}, \href{http://arxiv.org/abs/astro-ph/9911177}{{\ttfamily
  arXiv:astro-ph/9911177}}.

\bibitem{Planck:2015mym}
{\bfseries Planck} Collaboration, P.~A.~R. Ade {\em et~al.}, ``{Planck 2015
  results. XV. Gravitational lensing},''
  \href{http://dx.doi.org/10.1051/0004-6361/201525941}{{\em Astron. Astrophys.}
  {\bfseries 594} (2016) A15},
  \href{http://arxiv.org/abs/1502.01591}{{\ttfamily arXiv:1502.01591
  [astro-ph.CO]}}.

\bibitem{Lewis:2019xzd}
A.~Lewis, ``{GetDist: a Python package for analysing Monte Carlo samples},''
  \href{http://arxiv.org/abs/1910.13970}{{\ttfamily arXiv:1910.13970
  [astro-ph.IM]}}.

\bibitem{Assassi:2015fma}
V.~Assassi, D.~Baumann, and F.~Schmidt, ``{Galaxy Bias and Primordial
  Non-Gaussianity},''
  \href{http://dx.doi.org/10.1088/1475-7516/2015/12/043}{{\em JCAP} {\bfseries
  12} (2015) 043}, \href{http://arxiv.org/abs/1510.03723}{{\ttfamily
  arXiv:1510.03723 [astro-ph.CO]}}.

\bibitem{Villaescusa-Navarro:2013pva}
F.~Villaescusa-Navarro, F.~Marulli, M.~Viel, E.~Branchini, E.~Castorina,
  E.~Sefusatti, and S.~Saito, ``{Cosmology with massive neutrinos I: towards a
  realistic modeling of the relation between matter, haloes and galaxies},''
  \href{http://dx.doi.org/10.1088/1475-7516/2014/03/011}{{\em JCAP} {\bfseries
  03} (2014) 011}, \href{http://arxiv.org/abs/1311.0866}{{\ttfamily
  arXiv:1311.0866 [astro-ph.CO]}}.

\bibitem{Castorina:2013wga}
E.~Castorina, E.~Sefusatti, R.~K. Sheth, F.~Villaescusa-Navarro, and M.~Viel,
  ``{Cosmology with massive neutrinos II: on the universality of the halo mass
  function and bias},''
  \href{http://dx.doi.org/10.1088/1475-7516/2014/02/049}{{\em JCAP} {\bfseries
  02} (2014) 049}, \href{http://arxiv.org/abs/1311.1212}{{\ttfamily
  arXiv:1311.1212 [astro-ph.CO]}}.

\bibitem{Lazeyras:2015lgp}
T.~Lazeyras, C.~Wagner, T.~Baldauf, and F.~Schmidt, ``{Precision measurement of
  the local bias of dark matter halos},''
  \href{http://dx.doi.org/10.1088/1475-7516/2016/02/018}{{\em JCAP} {\bfseries
  02} (2016) 018}, \href{http://arxiv.org/abs/1511.01096}{{\ttfamily
  arXiv:1511.01096 [astro-ph.CO]}}.

\bibitem{jeong2010cosmology}
D.~Jeong, ``Cosmology with high (z> 1) redshift galaxy surveys,'' {\em Ph. D.
  Thesis} (2010) .

\bibitem{Ma:2005rc}
Z.-M. Ma, W.~Hu, and D.~Huterer, ``{Effect of photometric redshift
  uncertainties on weak lensing tomography},''
  \href{http://dx.doi.org/10.1086/497068}{{\em Astrophys. J.} {\bfseries 636}
  (2005) 21--29}, \href{http://arxiv.org/abs/astro-ph/0506614}{{\ttfamily
  arXiv:astro-ph/0506614}}.

\bibitem{Behroozi:2011ju}
P.~S. Behroozi, R.~H. Wechsler, and H.-Y. Wu, ``{The Rockstar Phase-Space
  Temporal Halo Finder and the Velocity Offsets of Cluster Cores},''
  \href{http://dx.doi.org/10.1088/0004-637X/762/2/109}{{\em Astrophys. J.}
  {\bfseries 762} (2013) 109}, \href{http://arxiv.org/abs/1110.4372}{{\ttfamily
  arXiv:1110.4372 [astro-ph.CO]}}.

\bibitem{Hamaus2011}
N.~{Hamaus}, U.~{Seljak}, and V.~{Desjacques}, ``{Optimal constraints on local
  primordial non-Gaussianity from the two-point statistics of large-scale
  structure},'' \href{http://arxiv.org/abs/1104.2321}{{\ttfamily
  arXiv:1104.2321 [astro-ph.CO]}}.

\end{thebibliography}\endgroup
\nocite{*}

\end{document}